\documentclass[aps,prd,reprint,superscriptaddress,floatfix]{revtex4-2}
\usepackage{graphicx}
\usepackage{amsmath}
\usepackage{url}
\usepackage[colorlinks]{hyperref}
\usepackage{orcidlink}
\usepackage{acronym}
\usepackage{xspace}

\newcommand{\past}{\ensuremath{p_\mathrm{astro}}}
\newcommand{\pter}{\ensuremath{p_\mathrm{terr}}}

\def\apjl{ApJL}

\begin{document}

\title{PyCBC Live search for compact binary mergers in Advanced LIGO and Virgo's fourth observing run}

\author{Max Trevor\,\orcidlink{0000-0002-2728-9508}}
\affiliation{University of Maryland, College Park, MD 20742, USA}
\email{mtrevor@umd.edu}

\author{Gareth S. {Cabourn Davies}\,\orcidlink{0000-0002-4289-3439}}
\affiliation{Institute for Cosmology and Gravitation, University of Portsmouth, 1-8 Burnaby Road, Portsmouth, P01 3FZ, UK}

\author{Tito {Dal Canton}\,\orcidlink{0000-0001-5078-9044}}
\affiliation{Universit\'e Paris-Saclay, CNRS/IN2P3, IJCLab, 91405 Orsay, France}

\author{Thomas Dent\,\orcidlink{0000-0003-1354-7809}}
\affiliation{Instituto Galego de F\'isica de Altas Enerx\'ias, Universidade de Santiago de Compostela, 15782 Santiago de Compostela, Galicia, Spain}

\author{Ian Harry\,\orcidlink{0000-0002-5304-9372}}
\affiliation{Institute for Cosmology and Gravitation, University of Portsmouth, 1-8 Burnaby Road, Portsmouth, P01 3FZ, UK}

\author{St\'ephanie Hoang}
\affiliation{Universit\'e Paris-Saclay, CNRS/IN2P3, IJCLab, 91405 Orsay, France}

\author{Arthur Tolley}
\affiliation{Institute for Cosmology and Gravitation, University of Portsmouth, 1-8 Burnaby Road, Portsmouth, P01 3FZ, UK}

\date{\today}

\begin{abstract}
PyCBC Live is a low-latency search pipeline that identifies gravitational waves from compact binary coalescences and provides alerts for electromagnetic follow-up. This paper presents improvements to PyCBC Live that were implemented for the fourth observing run (O4) of the LIGO-Virgo-KAGRA network, which operated from May 2023 to November 2025. The ranking statistic was enhanced to incorporate time-dependent background modeling using data quality streams and daily updates of the noise model. Follow-up capabilities were improved through refined probability of astrophysical origin calculations, optimized SNR recovery with reduced computational cost, and a method to incorporate Virgo as a sky-map-only detector. An Early Warning search was implemented to detect binary neutron star and neutron star-black hole systems before merger, providing pre-merger alerts with a pipeline latency of 2.5--3.5 seconds and warning times up to 60 seconds before coalescence. The autogating procedure was extended to apply to the full strain buffer rather than individual analysis segments, improving rejection of loud and rapidly successive glitches. Performance validation using the Mock Data Challenge showed sensitivity improvements of factors of 1.7 to 2.3 for the coincident search depending on source mass at an inverse false alarm rate of 10 years, and factors of 1.3 to 1.7 for the single-detector search. For injections in two-detector time, the O4 configuration identified 1979 of 2495 injections with a decisive SNR greater than 6 at a false alarm rate below one per year ($79.3\%$), compared to 1262 ($50.6\%$) with the O3 configuration. For injections in single-detector time, the O4 configuration identified 218 of 1174 injections ($18.6\%$), compared to 170 ($14.5\%$) with the O3 configuration. The search maintained a median latency of 15.94 seconds from merger to candidate upload.

\end{abstract}

\maketitle

\acrodef{GW}{gravitational wave}
\newcommand{\GW}{\ac{GW}\xspace}
\newcommand{\GWs}{\acp{GW}\xspace}
\acrodef{BNS}{binary neutron star}
\newcommand{\BNS}{\ac{BNS}\xspace}
\acrodef{BBH}{binary black hole}
\newcommand{\BBH}{\ac{BBH}\xspace}
\acrodef{NSBH}{neutron--star--black--hole}
\newcommand{\NSBH}{\ac{NSBH}\xspace}
\acrodef{CBC}{compact binary coalescence}
\newcommand{\CBC}{\ac{CBC}\xspace}
\newcommand{\CBCs}{\acp{CBC}\xspace}
\acrodef{FAR}{false-alarm rate}
\newcommand{\FAR}{\ac{FAR}\xspace}
\acrodef{IFAR}{inverse false-alarm rate}
\newcommand{\IFAR}{\ac{IFAR}\xspace}
\acrodef{EW}{early-warning}
\newcommand{\EW}{\ac{EW}\xspace}
\acrodef{SNR}{signal-to-noise ratio}
\newcommand{\SNR}{\ac{SNR}\xspace}
\newcommand{\SNRs}{\acp{SNR}\xspace}

\section{Introduction} \label{introduction}

The detection of \GWs from \CBCs has become routine since the first observation of GW150914~\cite{LIGOScientific:2016vbw}. The Advanced LIGO~\cite{LIGOScientific:2014pky}, Advanced Virgo~\cite{VIRGO:2014yos}, and KAGRA~\cite{KAGRA:2020tym, Somiya:2011np, Aso:2013eba} detector network identified multiple events per week during the fourth observing run (O4). The most recent catalog release, GWTC-5.0~\cite{LIGOScientific:2026wfs}, lists hundreds of \CBC signals.

Low-latency detection and characterization of these signals enables electromagnetic and neutrino follow-up observations, as demonstrated by the multi-messenger observations of the \BNS merger GW170817~\cite{GW170817MMA}. Rapid identification of \GW candidates requires analysis pipelines that balance high sensitivity with low latency. The pipelines used for rapid detection in O4 included cWB~\cite{Mishra:2024zzs}, GstLAL~\cite{Ewing:2023qqe}, MBTA~\cite{Allene:2025saz}, PyCBC Live~\cite{PyCBCLiveO2,PyCBCLiveO3}, and SPIIR~\cite{Chu:2020pjv}.

PyCBC Live was first deployed during the second observing run (O2) of Advanced LIGO~\cite{PyCBCLiveO2}. The pipeline identified GW170104 and GW170608 during non-nominal detector operation and provided the initial three-detector analysis of GW170817, producing a 31 deg$^2$ sky localization~\cite{Singer:2015ema} used to identify the optical counterpart AT 2017gfo. During the third observing run (O3), PyCBC Live continued operations with updated ranking statistics and improved follow-up capabilities~\cite{PyCBCLiveO3}.

The sensitivity of the LIGO detectors improved substantially in O4, with LIGO Hanford and LIGO Livingston achieving \BNS ranges exceeding 160 Mpc~\cite{Capote:2024rmo}. Virgo operated with comparable sensitivity to O3~\cite{Virgo:2019juy}. This sensitivity configuration, with two highly sensitive LIGO detectors and one less sensitive detector, motivated several updates to the PyCBC Live search strategy. 

O4 was divided into three parts: O4a (May 2023 -- January 2024), O4b (April 2024 -- January 2025), and O4c (January -- November 2025); some of the improvements described in this paper were deployed progressively across these periods. PyCBC Live's first detection of the O4 era was GW230518\_125908, a potential \NSBH merger that occurred during an engineering run six days before the start of O4 \cite{LIGOScientific:2025slb}. Over the course of O4, PyCBC Live contributed to the low-latency detections of numerous high-profile events~\cite{LIGOScientific:2025hdt, LIGOScientific:2026wfs}, including GW230529\_181500~\cite{LIGOScientific:2024elc}, GW231123\_135430~\cite{LIGOScientific:2025rsn}, GW241011\_233834, GW241110\_124123~\cite{LIGOScientific:2025brd}, and GW250114\_082203~\cite{LIGOScientific:2025rid}.

This paper describes low-latency operations in Section~\ref{lowlatency} and improvements to PyCBC Live implemented for O4 in Section~\ref{Improvements}, including a new ranking statistic with time-dependent noise modeling, updated probability of astrophysical origin calculations, an early warning search for \BNS and \NSBH systems, \SNR optimization, the use of Virgo as a localization-only detector, and improved autogating. Section~\ref{performance} presents performance validation using the Mock Data Challenge.

\section{Low latency operations} \label{lowlatency}

The full set of low-latency data products is described in~\cite{Chaudhary:2023vec}; we summarize the relevant details here. Search pipelines like PyCBC performed real-time matched filtering of the detector strain using a bank of simulated signal waveforms called \textit{templates}. High peaks in the matched-filter \SNR were identified and stored as \textit{triggers}. Both individual triggers and coincident trigger pairs were assigned a \FAR. Low-latency alerts were sent for events with \FAR below $2.3 \times 10^{-5}$\,Hz (two per day); these were deemed low-significance candidates. Events with \FAR below a lower threshold were considered to be significant. The significant \FAR threshold changed over the course of O4. It was always set as 1 per N months, with N being a trials factor based on the number of active low-latency CBC searches. This was chosen so that the combined rate of significant false alarms from all CBC searches would be one per month. 

When a pipeline identified a candidate below the low-significance \FAR threshold, it was uploaded to GraceDB~\cite{Chaudhary:2023vec}, a database that aggregates event candidates from all search pipelines and triggers automated follow-up, and a machine-readable notice was distributed via the General Coordinates Network (GCN). Any subsequent events with a coalescence time within 1 second of the first upload were grouped into the same superevent. BAYESTAR~\cite{Singer:2015ema} automatically computed spatial localizations for each uploaded event. After a delay of $\mathcal{O}(\mathrm{minutes})$ to collect uploads from all pipelines, a preferred event was selected from the superevent and an additional notice was sent . Notices included the \FAR of the preferred event, and inferred source properties including the spatial location, probability of astrophysical origin and source classification. Significant events underwent human review before a human-readable GCN circular was issued.

O4 introduced \EW alerts~\cite{Nitz:2020vym, Magee:2021xdx}, which could be sent before the coalescence of a \BNS or \NSBH system. These used specialized search methods to identify the inspiral signal before merger, enabling astronomers to begin searching for an electromagnetic counterpart as early as possible.

\section{Method Improvements for O4}
\label{Improvements}
This section describes the improvements made to PyCBC Live for O4. The resulting sensitivity gains are quantified in Sec.~\ref{performance} using the Mock Data Challenge (MDC).

\subsection{Low-latency data quality information} \label{dataquality}
Detector noise is non-Gaussian~\cite{LIGO:2024kkz}. Short bursts of excess broadband noise, \textit{glitches}, produce more high-\SNR triggers than expected from Gaussian noise. We refer to time segments containing glitches as \textit{glitchy}. Different glitch morphologies preferentially excite specific template subsets~\cite{Yarbrough:2025nsa}: for example, blip glitches resemble high-mass binary black hole waveforms and trigger the corresponding templates at elevated \SNR more often than other templates in the bank. To assess the impact of non-Gaussian noise in different regions of the search parameter space, we group templates into bins based on their template duration.

During O4, we used time series produced in low latency by the iDQ~\cite{Essick:2020qpo} data quality pipeline to identify time segments likely to contain glitches. Whenever a glitch was reported by iDQ with a false-alarm probability less than 0.001, a segment was constructed from one second before the glitch to one second after.

During O4a~\cite{LIGOScientific:2025slb} and O4b~\cite{LIGOScientific:2026wfs}, triggers falling into glitchy segments were vetoed to reduce the rate of false alarms. Starting in O4c, triggers falling within glitchy segments were instead marked as glitchy. These segments and triggers were used to calculate time-dependent trigger rate correction factors, which were applied to glitchy triggers when calculating the ranking statistic described in the next section.

\subsection{Updated ranking statistic} \label{ranking}
The ranking statistic assigns a significance to each \GW candidate and maps directly to \FAR. In O3, the ranking statistic was constructed from the probability $p_S\left(\vec{\theta}\right)$ of observed time and phase differences  between detectors and a quadrature sum of the reweighted network \SNR~\cite{PyCBCLiveO3}:
\begin{equation}
R^2 = \hat{\rho}_H^2 + \hat{\rho}_L^2 + 2\ln\left(p_S(\vec{\theta})\right).
\label{quad_sum_stat}
\end{equation}
where $\hat{\rho}_{(H/L)}$ are the reweighted \SNRs from the LIGO Hanford and LIGO Livingston observatories respectively.
The reweighted \SNR $\hat{\rho}$ is derived from the matched-filter \SNR using a $\chi^2$ signal consistency test~\cite{Allen:2004gu}. At the start of O4 we extended the definition of $\hat{\rho}$ to also include a high-frequency sine-Gaussian discriminator~\cite{Nitz:2017lco}.

In O4c, we adopted a ranking statistic developed for the offline search~\cite{Davis:2022cmw}, which addressed template-dependent noise behavior in two ways: by fitting the trigger rate exponentially for each template, and by applying time-dependent trigger rate corrections for each template bin.

The improved ranking statistic is defined as the logarithm of the likelihood ratio:
\begin{equation}
R(\vec{\kappa}) = \log r_S(\vec{\kappa}) - \log r_N(\vec{\kappa}),
\end{equation}
where $\vec{\kappa}$ represents the trigger properties, including the template parameters $\vec{\theta}$ and the trigger times in each detector, and $r_S$ and $r_N$ are the rates of astrophysical signals and noise events respectively. The signal model $r_S$ follows the O3 offline search~\cite{Davies:2020tsx}. The coincident noise model $r_N$ is proportional to the product of the single-detector noise models $r_{N,d}$~\cite{Nitz:2017svb}:
\begin{equation}
r_{N,d}(\hat{\rho}; \vec{\theta}; t) = \mu(\vec{\theta})\delta(\vec{\theta}, t)\alpha(\vec{\theta}) \exp[-\alpha(\vec{\theta})(\hat{\rho} - \hat{\rho}_{\text{th}})].
\end{equation}
The constituent parts of the single-detector noise model are:
\begin{itemize}
    \item $\mu(\vec{\theta})$, the rate of triggers with reweighted \SNR $\hat{\rho}$ greater than $\hat{\rho}_{\text{th}}$ in template $\vec{\theta}$
    \item $\alpha(\vec{\theta})$, the exponential decay constant
    \item $\delta(\vec{\theta}, t)$, the time-dependent correction factor \cite{Davis:2022cmw}.
\end{itemize}

In the offline PyCBC search, data are analyzed in chunks of a few weeks, and the noise model for each chunk is based on all triggers within it. For the low-latency analysis there are no predefined chunks, so we reconstruct the noise model each day using triggers from the previous days.  We include only triggers with $\hat{\rho} > \hat{\rho}_{\text{th}}$ in the daily fitting process. In O4, the previous 20 days of data were used to fit the noise model, and we used a threshold $\hat{\rho}_{\text{th}}=6$.

For each template, we model the rate of triggers as a decreasing exponential function of $\hat{\rho}$ before applying the time-dependent correction. Maximum-likelihood estimation is used to fit the $\mu$ and $\alpha$ parameters, which are then smoothed over nearby templates.

We then calculate the time-dependent correction factor $\delta$ for each template bin. The original offline iDQ implementation \cite{Davis:2022cmw} used 100 time bins defined by percentiles of the iDQ log-likelihood; for low-latency use we simplify this to 2 bins corresponding to glitchy and clean (non-glitchy) segments. The use of a smaller number of time bins avoids small sample sizes when calculating $\delta$ from a limited amount of background data.

For each template bin, the correction factor for glitchy segments is calculated by dividing the trigger rate during glitchy segments by the overall trigger rate in that bin:
\begin{equation}
	\delta = \frac{N_{\text{glitchy}}}{T_{\text{glitchy}}} \left( \frac{N_{\text{total}}}{T_{\text{total}}} \right)^{-1}
\end{equation}
An analogous correction factor is calculated for clean segments. The results of these model fits are used to calculate the ranking statistic $R$ for pairs of coincident triggers. The ranking statistic is mapped to \FAR using time-slides~\cite{Usman:2015kfa}.

PyCBC Live is also able to search for single-detector events, following the method developed for the offline search~\cite{Davies:2022thw}. In O4, the single-detector search was restricted to \BNS and \NSBH systems expected to produce electromagnetic counterparts.

The single-detector ranking statistic is obtained from the coincident statistic by dropping the terms that depend on trigger coincidence, giving
\begin{equation}
R = -\log r_{N}(\hat{\rho}; \vec{\theta}; t) + R_\sigma(\vec{\theta}),
\end{equation}
where $R_\sigma$ is the logarithm of the sensitive volume for template $\vec{\theta}$ normalized by a reference volume \cite{Davies:2022thw}.

To map ranking statistic to \FAR for single-detector triggers, we sort the relevant templates into template bins based on their duration. In O4, the set of bins used for this calculation was different from the bins used to calculate the time-dependent correction factors described in the previous section.
We only calculate \FAR for single detector triggers above a ranking stat threshold $R_{\mathrm{thresh} }$. The \FAR in detector $d$ and template bin $i$ is modeled as a decreasing exponential function of the ranking statistic:
\begin{equation}
\text{FAR}_{di}(R) = N_{d,i}\alpha_{d,i} \exp[-\alpha_{d,i}(R - R_{\text{thresh}})].
\end{equation}
where N is the total rate of triggers above 
In O4, this model was refit each day using single-detector triggers from the previous 20 days.

\subsection{Probability of astrophysical origin and source classification} \label{pastro}

Beyond the \FAR, PyCBC also estimates the probability of astrophysical origin $\past$, and its complement, the probability of terrestrial origin, $\pter$. These estimates incorporate knowledge of the rate of astrophysical mergers from previous observing runs. Let $\Re_N(\vec{\kappa})$ and $\Re_S(\vec{\kappa})$ denote the rate densities of noise and signal events as functions of the trigger properties $\vec{\kappa}$. The probability of astrophysical origin is then~\cite{LIGOScientific:2016kwr}
\begin{equation}
    \frac{\past}{1 - \past} \equiv \frac{\past}{\pter} = \frac{\Re_S(\vec{\kappa})} {\Re_N(\vec{\kappa})}.
\end{equation}
These densities vary over the trigger parameter space. For the online calculation, we approximate the key dependencies using information available from the search, focusing on:
\begin{itemize}
    \item The variation of signal and noise densities over the bank parameters (binary masses and spins), denoted schematically as $\vec{\theta}$.
    \item The variation of densities over a trigger statistic $\rho_c$, which we take to be \SNR-like, i.e.\ approximately proportional to the quadrature sum of single-detector \SNRs up to $\chi^2$ and network consistency corrections. In O4 we used the network SNR $\rho_c^2 = \rho_H^2 + \rho_L^2$
\end{itemize}
We factor the total rate density into a rate of triggers above threshold in each region of the bank, $r_{N}(\vec{\theta})$ or $r_{S}(\vec{\theta})$, multiplied by a normalized PDF over $\rho_c$:
\begin{equation}
    \Re_{S,N} = r_{S,N}(\vec{\theta}) f_{S,N}(\rho_c|\vec{\theta}).
\end{equation}

\subsubsection{Modelling event PDFs over ranking statistic}

We model the background PDF with an exponential in $\rho_c$ following~\cite{Nitz:2017svb}:
\begin{equation}
    f_N(\rho_c|\theta) \simeq \alpha_c \exp(-\alpha_c (\rho_c - \rho_\mathrm{th})) \quad (\rho_c > \rho_\mathrm{th}),
\end{equation}
where $\alpha_c$ is a slope parameter (see also~\cite{Lynch:2018yom}). For lower-mass templates, which correspond to the systems most relevant for electromagnetic follow-up, we find $\alpha_c = 6$ is a good approximation of the distribution. We also use this value for higher-mass systems: although the corresponding templates have noise distributions with longer tails, so this choice yields a conservatively high (i.e.\ pessimistic) estimate of $f_N$.

We model the signal PDF using the ideal \SNR distribution $p(\rho_c) \sim \rho_c^{-4}$~\cite{Schutz:2011tw}, taken to be independent of the template parameters:
\begin{equation}
    f_S(\rho_c) \simeq 3 \rho_\mathrm{th}^3 \rho_c^{-4}  \quad (\rho_c > \rho_\mathrm{th}),
\end{equation}
where $\rho$ is the network \SNR recovered by the search.

\subsubsection{Template dependence}

For the O3 \past{} estimate, the template bank was divided into 3 chirp mass bins~\cite{LIGOScientific:2021usb}, which provided only a coarse approximation of the signal distribution. For O4, we increased the number of bins (indexed by $p$) to 8, each with a signal trigger rate $r_{S,p}$ and noise trigger rate $r_{N,p}$.  The signal rate in each bin is estimated from O1--O3 detections, scaled by a factor accounting for the sensitivity of the active detector network relative to O3.  In the O4 configuration, bins were spaced by powers of 2 in template chirp mass, except for the highest bin.

For the background, the \FAR at a given $\rho_c$ is the integral of $\Re_N$ above $\rho_c$. With the exponential model above, the rate \emph{density} at $\rho_c$ equals $\alpha_c \times \mathrm{FAR}(\rho_c)$. Taking the noise trigger rate to be proportional to the template count $n^t_p$ in each bin, the noise rate density in bin $p$ is
\begin{equation}
    \Re_{N,p}(\rho_c) = \alpha_c \cdot \mathrm{FAR(\rho_c)} \frac{n^t_p}{\sum_p n^t_p}. 
\end{equation} 

For the signal rate, we estimate network sensitivity via the \BNS inspiral horizon $d_{h,D}$ for each detector $D$ (approximately $2.26$ times the inspiral range). Since the \SNR of a signal at distance $d_L$ is proportional to $d_{h,D}$, the network \SNR scales as 
\begin{equation}
    \rho_c^2 \propto d_L^{-2} \sum_{D} d_{h,D}^2 . 
\end{equation}
The rate of signals detected above a \SNR threshold $\rho_\mathrm{th}$ scales as the cube of the distance corresponding to a given network \SNR, i.e.\
\begin{equation}
 r_S \propto \left( \rho_\mathrm{th}^{-2} \sum_D d_{h,D}^2 \right)^{3/2}.
\end{equation}
Thus, we approximate the total rate of signals above \SNR threshold in bin $p$ as
\begin{equation}
    r_{S,p} = r_{p,\mathrm{O3}} \left( \sum_D d_{h,D}^2 \right)^{3/2}
    \left( \sum_D d_{h,D(\mathrm{O3})}^2 \right)^{-3/2},
\end{equation}
where O3 quantities $r_{p,\mathrm{O3}}$ and $d_{h,D(\mathrm{O3})}$ are the O3 detection counts and representative horizon distances respectively. For reference, \BNS ranges of $[135, 110, 50]$\,Mpc for LIGO Livingston Observatory (LLO), LIGO Hanford Observatory (LHO), and Virgo give a network horizon distance $[\sum_D d_{h,D(\mathrm{O3})}^2]^{1/2} = 410$\,Mpc.

\begin{figure}[!t]
    \centering
    \includegraphics[width=0.8\columnwidth]{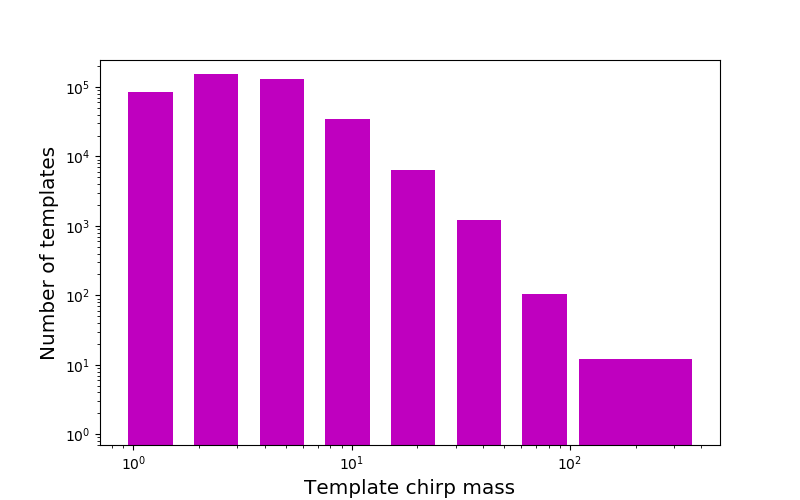}
    \includegraphics[width=0.8\columnwidth]{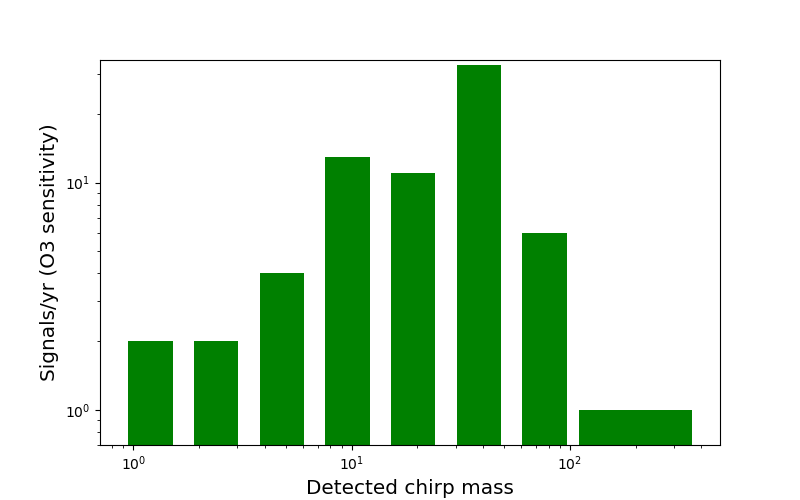}
    \caption{Template counts (top) and O1--O3 detection counts (bottom) per chirp mass bin ($M_\odot$), used to compute the $\past$ estimates.}
    \label{fig:template_signal_hists}
\end{figure}
We show in Fig.~\ref{fig:template_signal_hists} the binned template counts for the O4 PyCBC Live template bank, and previous detection counts for observations up to O3.  We then have 
\begin{equation}\label{eq:mainapprox}
    \frac{\past}{\pter} \simeq 
    3 \frac{\rho^3_{\rm th}}{\rho_c^4}
    \frac{r_{S,p} \sum_p n^t_p}{\alpha_c \cdot \mathrm{FAR}(\rho_c) n^t_p} .
\end{equation}

This relation requires minor corrections for practical implementation. The \FAR estimate has a finite precision limited by the available background time: for two-detector coincidences the minimum \FAR is 1 per 100 years. For loud candidates that saturate this limit, the $\rho_c$ dependence of Eq.~\eqref{eq:mainapprox} is modified as described in Sec.~4.2 of~\cite{o4_live_pastro_techdoc}. In O3, multiple detector combinations were considered for detection~\cite{PyCBCLiveO3}, requiring separate \FAR estimates and signal rates for each event type; in O4 only LHO-LLO coincidences and single-detector events were used, simplifying the accounting (see Sec.~4.1 of~\cite{o4_live_pastro_techdoc}). 

\subsubsection{Binary source type classification} 

The resulting \past{} value covers all astrophysical compact binary source classes (\BNS, \BBH, and \NSBH). For electromagnetic follow-up planning, the relative probabilities of these classes are also needed. In PyCBC Live, these are estimated from the template chirp mass and a redshift estimate based on the trigger \SNRs and detector horizon distances for the trigger template ~\cite{Villa-Ortega:2022qdo}. The relative class probabilities are then multiplied by \past{} to obtain absolute class probabilities, which are reported in public alerts together with $\pter \equiv 1 - \past{}$~\cite{userguide-inference}.

\subsection{Early warning search} \label{earlywarning}

Because the lower frequency limit of LIGO-like detectors is $\sim$20 Hz, a $1.4 + 1.4\,M_\odot$ \BNS system can exist in the detector sensitive band for $\sim 160$ s before coalescence, and this time would be even longer for lower-mass systems. Potential electromagnetic counterparts, like a kilonova \cite{Metzger:2019zeh} or short gamma-ray burst \cite{Burns:2019byj}, are most likely produced at or after merger, and typically rapidly fade on time scales of seconds to days. Therefore, detecting the inspiral signal and localizing the source before coalescence provides an important warning window during which telescopes can be slewed to the approximate sky location and increase the chance of observing the short-lived counterpart \cite{MergerForecasting, SwiftlyChasing}. Since O4, PyCBC Live uses a dedicated \emph{\EW} search to issue such pre-merger alerts.

The \EW search uses frequency-truncated templates: waveforms that begin at the lower frequency of the LIGO band and end at an upper cutoff frequency rather than continuing through merger \cite{MergerForecasting}.
By contrast, we refer to the post-merger search, which uses non-truncated templates, as the \emph{full-bandwidth} search.
Six cutoff frequencies were used in O4—29, 32, 38, 44, 49, and 56 Hz—corresponding to approximately 60, 46, 29, 20, 15, and 10 seconds before merger for a $1.4 + 1.4\,M_\odot$ \BNS. At the 29 Hz cutoff, only $\sim$31\% of the total matched-filter \SNR has been accumulated; the remaining \SNR builds up as the signal sweeps to higher frequencies.

The O4 \EW template bank was constructed as six separate sub-banks (one per cutoff) using a geometric placement algorithm \cite{Brown:2012qf} with a minimum match of 0.97 and a noise power spectral density representative of the Advanced LIGO design sensitivity with a \BNS range of 140 Mpc \cite{aLIGODesign}. Templates covered component masses of $1$–$3\,M_\odot$ with zero spin, targeting electromagnetically bright \BNS and \NSBH systems. The six sub-banks contained a total of 9180 templates, compared to $\sim$700,000 in the full-bandwidth search \cite{Tolley:2025}.

The \EW search is designed to minimize the time an alert takes to reach astronomers. The pipeline latency—time from the end of a data segment to candidate upload—has three components: the 1-second analysis stride (data must accumulate before processing), a 1.5-second overwhitening padding period required to avoid data conditioning artifacts at the edges of each segment, and the matched-filter processing time, which must complete within one stride. This yields a total latency of 2.5–3.5 seconds, compared to 12--20 seconds for the full-bandwidth search. To ensure processing fits within this budget, the \EW search restricts matched filtering to the two LIGO detectors (LIGO-Hanford and LIGO-Livingston) and uses a simplified ranking statistic: a single-detector \SNR threshold combined with the phase-time-amplitude coincident histogram statistic (Equation~\ref{quad_sum_stat}), rather than the more computationally expensive statistic used by the full-bandwidth search \cite{Tolley:2025}.

Spatial localization is provided by BAYESTAR \cite{Singer:2015ema}, which can produce results in seconds from matched-filter trigger information. Because \EW candidates are detected before the full inspiral \SNR has accumulated, early localizations are coarse but improve systematically as the signal is detected at successively higher frequency cutoffs with increasing accumulated \SNR. Adding a third detector to the network would substantially improve localization at all cutoff frequencies, but Virgo is not used by the \EW search to maintain the strict latency requirement.

For a GW170817-like injection in O4 noise~\cite{Capote:2024rmo} placed at a distance giving a full-bandwidth network \SNR of 30, the 90\% credible sky area decreases from $\sim$16,000 deg$^2$ at 29 Hz to $\sim$2,300 deg$^2$ at 56 Hz, approaching 149 deg$^2$ at full bandwidth \cite{Tolley:2025}. The 29 Hz template detects the signal 67 seconds before merger. The original GW170817 detection alert was issued 27 minutes post-merger, largely due to a loud glitch in LIGO-Livingston that required manual data conditioning before a localization could be produced \cite{GW170817}. An \EW search operating during a GW170817-like event today would provide up to a minute of pre-merger warning, giving electromagnetic observatories time to begin slewing before the merger and counterpart occur.

\subsection{SNR Optimizer} \label{snropt}

The \SNR optimizer is a post-detection algorithm that refines the template parameters of a detected event via a series of additional matched-filter searches, recovering signal power lost due to the discrete spacing of the template bank \cite{PyCBCLiveO3}. The BAYESTAR~\cite{Singer:2015ema} sky localization algorithm ingests \SNR timeseries from each detector around the time of the event; higher \SNR improves sky localization accuracy. This makes \SNR optimization valuable for multi-messenger astronomy. The optimizer was used in O3 and in O4b and O4c, but was disabled during O4a due to technical issues.

The loss of some portion of the true \SNR is inevitable due to the discrete nature of template banks. PyCBC template banks are constructed with a minimum match criterion of 0.97, permitting up to $3\%$ \SNR loss for any \GW event. When a signal is detected, the \SNR optimizer performs a refined search over the template parameter space surrounding the initially detected template to recover this potentially lost \SNR. Following the initial event upload to GraceDB, PyCBC Live spawns a subprocess running the optimization algorithm, which maximizes network \SNR by tuning four template parameters: primary component mass, secondary component mass, and the z-components of both object spins. Parameter boundaries are calculated from the initial event values to constrain the search space.

The \SNR optimizer employs the differential evolution algorithm~\cite{Storn:1997} as implemented in SciPy~\cite{Virtanen:2019joe}.
The four key hyperparameters of the differential evolution algorithm are the population size $N$, maximum number of iterations, mutation factor, and recombination constant. Between O3 and O4 the population size was reduced from 200 to 100 and the maximum number of iterations from 100 to 50, reducing runtime by approximately 25\% while maintaining \SNR recovery performance.

Compared to the O3 implementation, the parameter space boundaries have been refined in two ways. First, chirp mass bounds are derived from a $\pm 2$\,s window in chirp time $\tau_0 \propto \mathcal{M}^{-5/3}$ around the initial event's chirp time, providing a more physically motivated search region than a flat chirp mass window. Second, astrophysically motivated spin bounds are applied based on the component masses: for templates whose maximum component mass falls below $3\,M_\odot$, spin magnitudes are constrained to $0.4$, consistent with the theoretical neutron star spin limit; for higher-mass templates, the bounds are relaxed to $0.9$. These tighter bounds avoid searching physically irrelevant regions of parameter space, reducing optimization time.

The performance of the \SNR optimizer for simulated signals is described in Sec.~\ref{snr-opt-performance}.

\subsection{Usage of Virgo for spatial localization} \label{virgo}

The PyCBC Live search used in O3 treated all three \GW detectors (LHO, LLO, and Virgo) equally \cite{PyCBCLiveO3}. In O4, the sensitivity of the two LIGO detectors was substantially increased while the sensitivity of Virgo was largely unchanged. This sensitivity difference increased the likelihood of a \GW being seen in only the Hanford and Livingston detectors. Treating all three detectors as equal increases the computational cost of the analysis, while at the same time introducing a relatively large trials factor in the \FAR calculation, which can be detrimental if one of the detectors in not as sensitive as the others.

After these considerations, for the O4 analysis, we treated Virgo as a \emph{follow-up} detector rather than a triggering detector, similarly to what was done in O2 \cite{PyCBCLiveO2}. We did not continuously perform matched filtering of the Virgo strain data. Instead, when a \GW was identified in one or both LIGO detectors with \FAR below the low-significance threshold, we performed matched filtering of the Virgo strain data with the reported template to produce an \SNR time series. This time series was then included in the upload to GraceDB and used in the spatial localization of the source \cite{Singer:2015ema}, achieving a more informative localization than would have been possible with two detectors alone.
With respect to the implementation used in O2, the ability to use certain detectors only for the spatial localization has been generalized, and is now applicable to KAGRA as well with a simple configuration change.

The \SNR optimizer used only the strain data from the LIGO detectors as inputs; however, we used the best-fit parameters reported by the optimizer to perform a further matched filter of the Virgo data. The \SNR time series of all three detectors using the optimized parameters were then uploaded to GraceDB and used to update the spatial localization.

\subsection{Improved autogating} \label{autogating}

Like many other signal detection algorithms employed on \GW data, PyCBC Live uses a data conditioning procedure called \emph{gating} in order to suppress large-amplitude glitches, which would damage the ability of the search to detect astrophysical signals if left in the data.
Gating consists in zeroing-out a short data segment around a glitch, typically for a maximum duration of a couple seconds, and tapering the zeroed-out segment to avoid discontinuities at the edges, which would themselves produce a similar effect as the glitch being removed \cite{Usman:2015kfa}.
Gating requires knowing the occurrence times of loud glitches.
In PyCBC-based searches, including PyCBC Live, these are usually determined via a simple glitch-finding algorithm that looks for large excursions of the magnitude of the whitened strain time series.
The resulting complete procedure, glitch identification from the strain data followed by gating, is referred to as \emph{autogating}.

In O3, PyCBC Live applied autogating only once per analysis segment, which has a duration of 8 s only \cite{PyCBCLiveO3}.
Although still beneficial compared to the absence of autogating, this approach was found to gate significantly fewer glitches that would otherwise be gated if a longer segment could be used.
The main reasons are that i) glitches near the edges of the analysis segment cannot be easily detected due to the impulse response of the involved filters, and ii) many glitches are actually composed of multiple short transients in rapid succession, sometimes covering a total duration of a few seconds, and the basic glitch detector can fail to detect all of them if they are too close in time.
This proved to be particularly problematic for the early-warning analysis, due to the analysis segment being even shorter than the full-bandwidth analysis.

For O4, we instead applied autogating to the entire rolling window of time-domain strain that goes into the FFT, which must be much longer than the analysis segment in order to accomodate the full length of the template and the impulse response of the various filters involved in the calculation of the \SNR.
This effectively gives autogating multiple chances to gate the same glitch, and multiple closely-spaced glitches can be progressively gated over multiple analysis segments.
The result is a much higher probability that even multiple loud glitches in rapid succession will be correctly gated at some point before they are shifted out of the strain buffer, enabling the detection of long CBC signals that overlap the earlier glitches.

\section{Performance} \label{performance}

\subsection{Test setup}
We compared the search sensitivity using the updated ranking statistic against a search  using the O3-era statistic. Comparisons were made using the Mock Data Challenge (MDC) described in~\cite{Chaudhary:2023}, which replayed 40 days of O3 data with injected simulated compact binary signals to test the low-latency alert infrastructure.

The injection set consisted of 50,000 simulated \GW signals spanning the full compact binary parameter space: $40.9\%$ \BNS systems, $35.8\%$ \NSBH systems, and $23.3\%$ \BBH systems. Component masses ranged from $1\,M_\odot$ for neutron stars up to $100\,M_\odot$ for black holes, with mass distributions following astrophysically motivated power laws~\cite{Chaudhary:2023}. Spins were distributed uniformly in magnitude with isotropic orientations, subject to theoretical constraints of a maximum neutron star spin magnitude of 0.4 and a maximum black hole spin magnitude of 0.998. The IMRPhenomPv2\_NRTidalv2 waveform approximant~\cite{Hannam:2013oca,Dietrich:2019kaq} was used to model matter effects in systems containing neutron stars, with the SLy equation of state~\cite{Douchin:2001sv} setting the neutron star maximum mass at approximately $2.05\,M_\odot$. Injected signals were distributed uniformly in comoving volume assuming flat $\mathrm{\Lambda CDM}$ cosmology  with $\mathrm{H}_0 = 67.3 \mathrm{km}~\mathrm{s}^{-1}~ \mathrm{Mpc}^{-1}$ and $\Omega_m = 0.3$, based on Planck 2018 results~\cite{Planck:2018vyg}. Maximum redshifts were tailored to source type: \BNS systems to redshift $0.15$, \NSBH systems to $0.25$, and \BBH systems to $1.9$.

To assess detection performance, we used the \textit{decisive \SNR}: for injections occurring during two-detector time, this is the second-highest single-detector \SNR, i.e.\ the \SNR of the less sensitive detector that must observe the signal for a coincident detection; for injections during single-detector time, it equals the \SNR in the active detector. Injections with a decisive \SNR below 6 had negligible detection probability and were excluded from efficiency calculations. Of the 50,000 injections, 4318 had a decisive \SNR of 6 or more. Since the single-detector search did not target \BBH signals, \BBH injections occurring during single-detector time were also excluded from the efficiency calculation, leaving 3672 injections.

Additionally, we investigated the performance of the \SNR optimizer using the triggers produced from the test of the O4 ranking statistic on the MDC. We compared \SNR optimized events against their unoptimized counterparts.

The \EW search was tested on a set of 32,599 \BNS and \NSBH injections in simulated Gaussian noise. We chose not to use the MDC to test the \EW search because the MDC did not contain enough loud \BNS or \NSBH injections.

\subsection{Results}

\begin{figure}[tbp]
    \centering
    \includegraphics[width=\columnwidth]{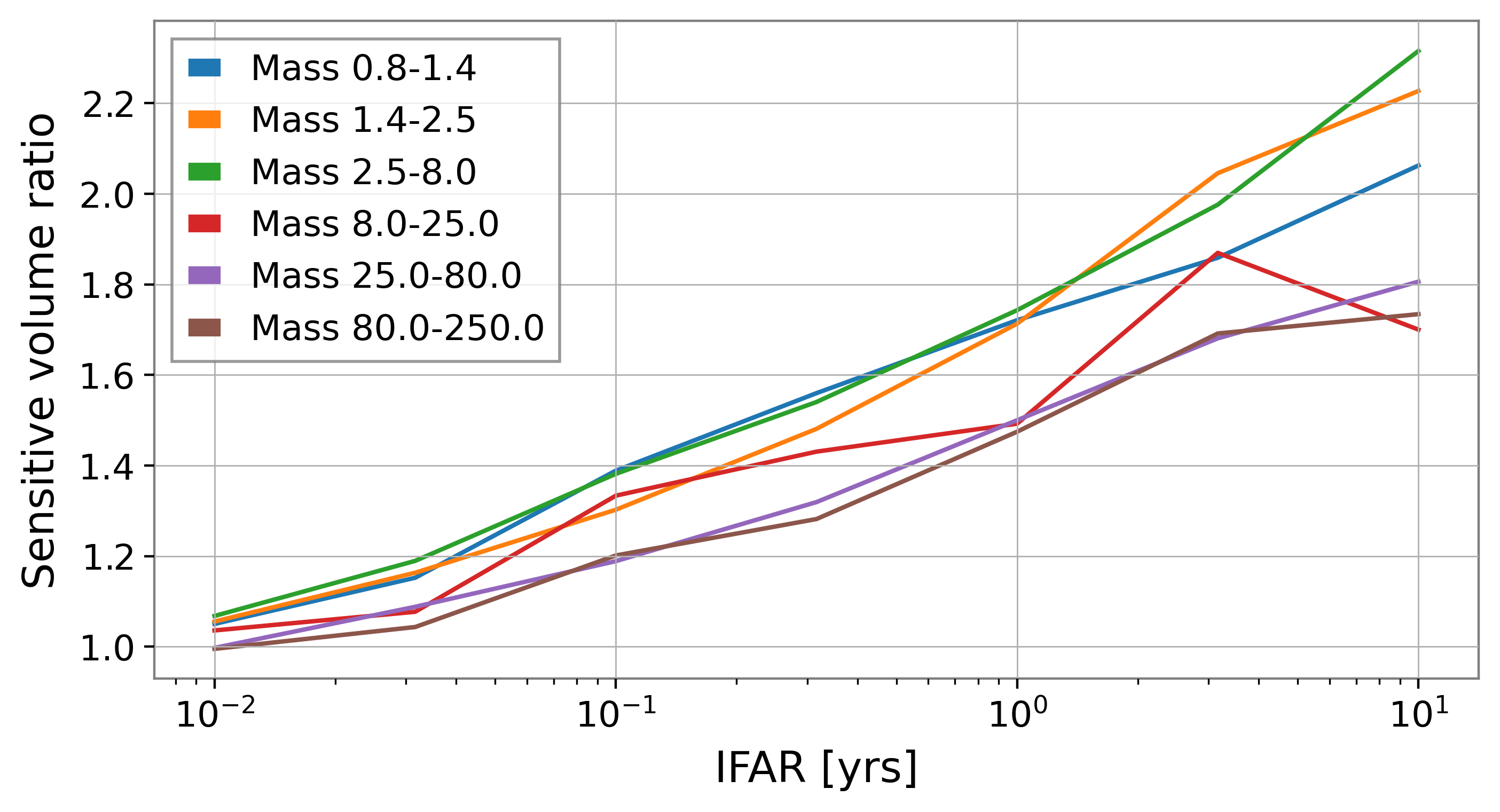}
    \includegraphics[width=\columnwidth]{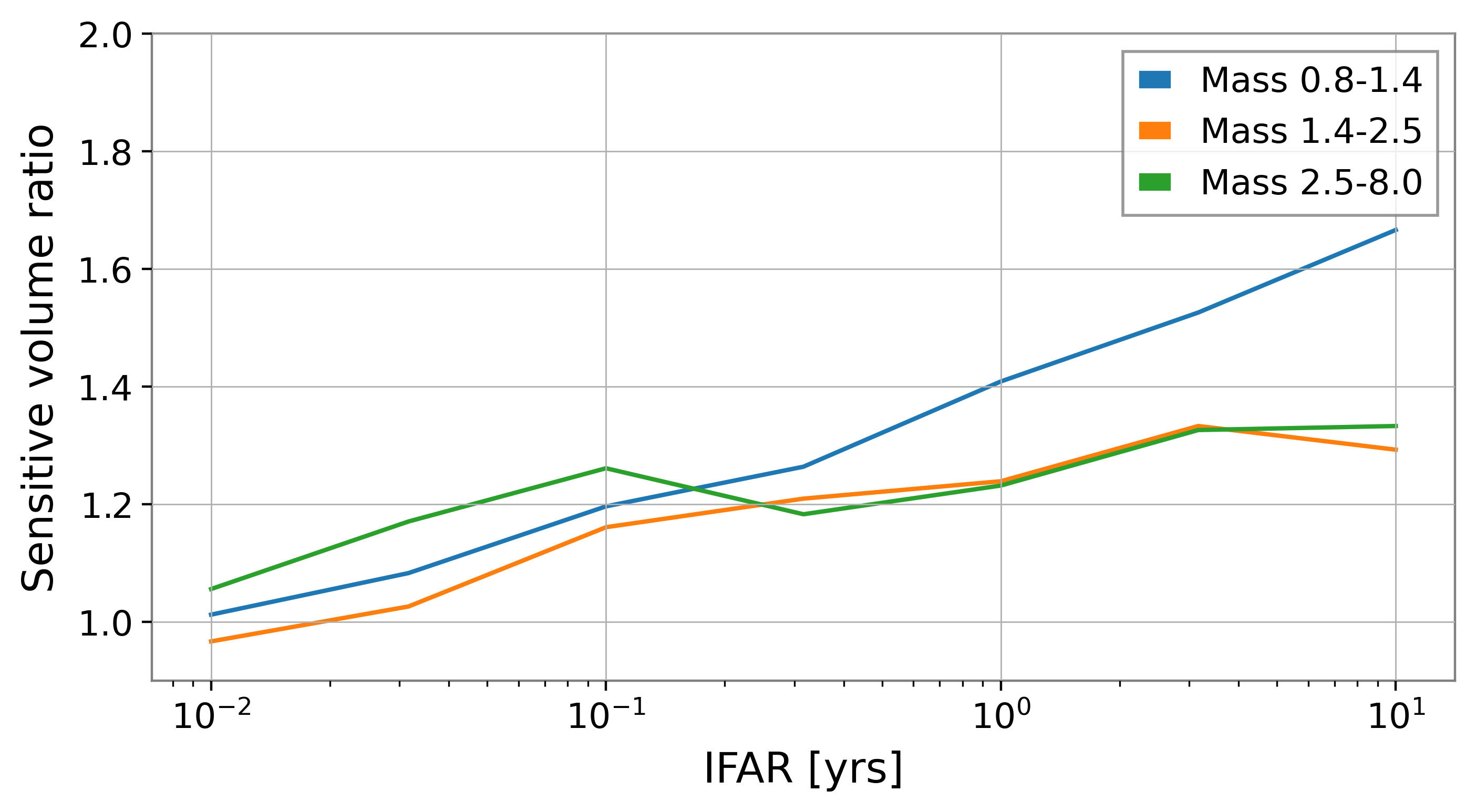}
    \caption{Sensitive volume ratio (O4 vs.\ O3 ranking statistic) as a function of \IFAR threshold; values above unity indicate improved sensitivity. (Top) Coincident search, for six chirp mass ranges ($M_\odot$). (Bottom) Single-detector \BNS and \NSBH search, for three chirp mass ranges ($M_\odot$).\label{fig:coincident_results}\label{fig:single_sensitivity}}
\end{figure}

\subsubsection{Search sensitivity}
\paragraph{Coincident search}

The updated ranking statistic described in this paper improved the sensitivity of the coincident search compared to O3.  The top of Figure~\ref{fig:coincident_results} shows the ratio of sensitive volume during two-detector time between the O4 and O3 ranking statistic configurations as a function of \IFAR threshold, for several source mass ranges.

At low \IFAR thresholds (below 0.1 years), all mass ranges converged to sensitive volume ratios near unity. At high \IFAR thresholds (10 years), the sensitive volume ratio depended strongly on source mass. Lower-mass systems (0.8--8.0~M$_\odot$) achieved the highest VT ratios, exceeding 2.0. The intermediate mass range (2.5--8.0~M$_\odot$) showed the largest sensitivity improvement. Higher-mass systems (8.0--250.0~M$_\odot$) showed smaller improvements, with VT ratios plateauing near 1.7.

The O4 configuration identified 1979 of the 2495 injections in two-detector time at a \FAR below one per year, an efficiency of $79.3\%$. The O3 configuration identified 1262 injections at the same threshold, an efficiency of $50.6\%$.

\paragraph{Single-Detector Search}

The performance of the single-detector search was measured using \BNS and \NSBH injections occurring during times when only one detector was active. The bottom of Figure~\ref{fig:single_sensitivity} shows the ratio of sensitive VT for the updated search compared to a single-detector search using the O3-era ranking statistic. The improvements were comparable to those seen in the coincident search, with VT ratios reaching approximately 1.3--1.7 at an \IFAR of 10 years, with the largest gains in the lowest mass bin. At lower \IFAR thresholds the ratios converged toward unity, a trend similar to that seen in the coincident search.

Of the 1174 \BNS and \NSBH injections with decisive \SNR above 6 occurring during single-detector time, the updated search identified 218 at a \FAR below one per year, an efficiency of $18.6\%$. The O3-era configuration identified 170 injections at the same threshold, an efficiency of $14.5\%$.

\subsubsection{Upload latency}

\begin{figure}[tbp]
    \centering
    \includegraphics[width=\columnwidth]{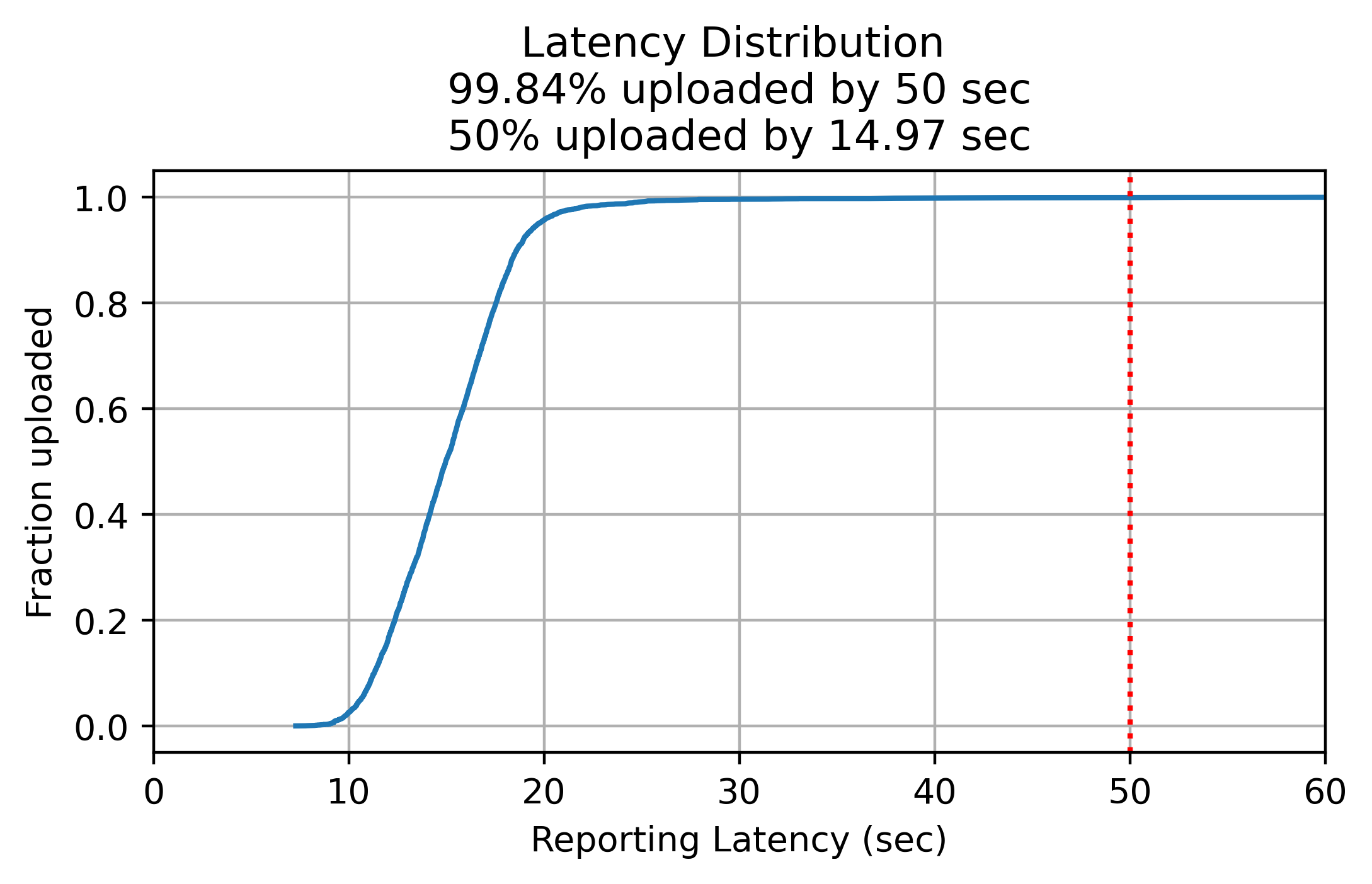}
    \caption{Cumulative distribution of merger-to-upload latencies on the MDC. Median: 15.94\,s; $99.89\%$ of events within 50\,s.}
    \label{fig:upload_latency}
\end{figure}

Figure~\ref{fig:upload_latency} shows the cumulative distribution of upload latencies for PyCBC Live events on the MDC. The median latency from merger to GraceDB upload was 15.94 seconds. LIGO operations imposed a requirement~\cite{Chaudhary:2023vec} that $99\%$ of uploads be completed within 50 seconds; PyCBC Live met this requirement, with $99.89\%$ of events uploaded within that threshold.

\subsubsection{SNR optimization} \label{snr-opt-performance}

\begin{figure}[!t]
    \centering
    \includegraphics[width=\columnwidth]{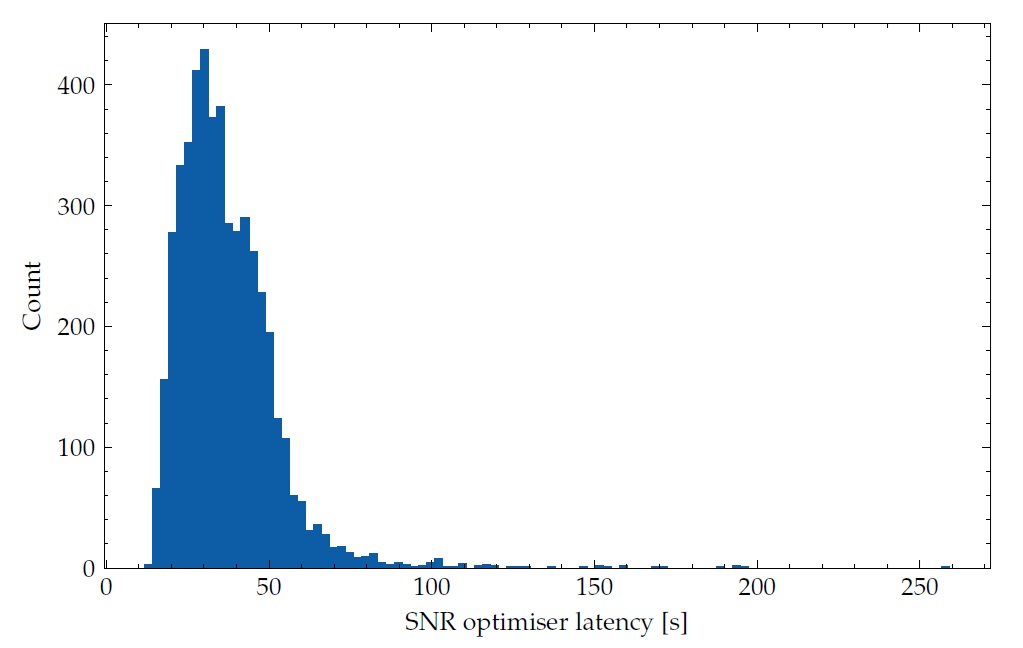}
    \includegraphics[width=\columnwidth]{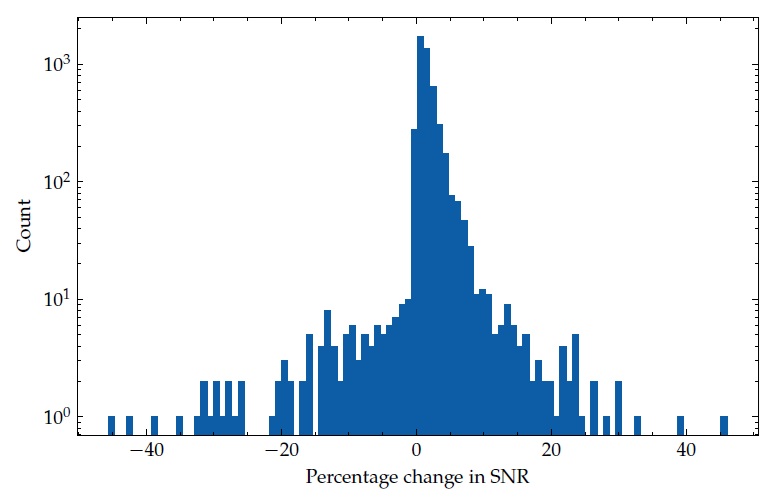}
    \caption{\SNR optimizer performance on the MDC~\cite{Tolley:2025}. (Top) Additional upload latency relative to the original PyCBC Live upload (mean: 37\,s). (Bottom) Network \SNR change after optimization (mean: $+1.64\%$).\label{fig:snr_opt_latency}\label{fig:snr_percent_change}}
\end{figure}

We evaluated the \SNR optimizer's performance on the MDC, retrieving all PyCBC full-bandwidth GraceDB events uploaded between 2024 May 30 and 2024 July 16~\cite{Tolley:2025}. Of 10,270 GraceDB events retrieved, 5,091 were uploaded by the \SNR optimizer. In 91.85\% of cases, the optimized event was preferred over the original PyCBC Live upload. The mean \SNR increase was $+1.64\%$, with a mean additional latency of 37 seconds. The distribution of SNR optimizer latency is shown in the top of Figure \ref{fig:snr_opt_latency}.

As shown in the bottom of Fig.~\ref{fig:snr_percent_change}, a minority of events ($8.15\%$) saw a decrease in \SNR after optimization. These cases occurred primarily for signals with high mass ratios, large precessing spins, or long template durations, where the optimizer's fixed number of iterations was insufficient to adequately explore the complex parameter space. Lower chirp mass events also required longer optimization times due to their extended template durations.

\subsubsection{Early warning search} \label{ew-performance}

The \EW search was evaluated using a dedicated injection campaign separate from the main MDC \cite{Tolley:2025}. A set of 32,599 simulated \BNS and \NSBH signals were injected into 40 days of simulated colored Gaussian noise, with component masses $1$--$3\,M_\odot$, near-zero spins (z-components drawn uniformly from $\pm 0.05$), and network \SNRs distributed uniformly between 12 and 60. Gaussian noise was used to isolate the effects of the search configuration from non-Gaussian transients, which are not suppressed by the simplified \EW ranking statistic.

Since each injection could in principle be detected at up to six frequency cutoffs, the maximum possible number of candidates was 195,594. The search found 175,219 total candidates. After removing 1,211 candidates with coalescence times inconsistent with any injection (pure false alarms) and 17,432 duplicate-frequency candidates caused by signals straddling analysis stride boundaries, 156,438 unique injection-associated candidates remained.

The main sensitivity problem identified was a population of missed candidates at individual frequency cutoffs, affecting 873 injections (2.68\%). The cause is that the \EW search uses a coincidence timing window of 13 ms (the 10 ms light-travel time between LIGO-Hanford and LIGO-Livingston plus a 3 ms margin for timing uncertainties), but the timing accuracy of frequency-truncated templates is limited by their small effective bandwidth and the 256 Hz sample rate of the \EW search. The 1$\sigma$ timing error ranges from 5.89 ms at 29 Hz to 1.89 ms at 56 Hz; at the lower frequency cutoffs this error can push coincident trigger pairs outside the allowed window. The solutions identified are to widen the coincidence window on a frequency-dependent basis, apply subsample interpolation to reduce sample-rate discretization errors, and regenerate the phase-time-amplitude ranking histogram with bins tuned to the 256 Hz sample spacing \cite{Tolley:2025}. The histogram could alternatively be replaced by a normalizing flow density estimator~\cite{Insley:2026tpg}, which has been shown to reduce storage requirements by over three orders of magnitude while maintaining sensitivity.

\section{Conclusions}

We have described the configuration and performance of PyCBC Live during the fourth observing run. The central improvement was a new ranking statistic incorporating template-dependent and time-dependent noise modeling. Per-template exponential noise models were fit daily to the preceding 20 days of triggers, and iDQ data quality streams were used to apply trigger rate corrections for glitchy time segments. This replaced the simpler O3 ranking statistic and was complemented by an updated $\past$ calculation using finer chirp mass binning and signal rates scaled from O1--O3 detections. Three follow-up capabilities were added or improved: an \SNR optimizer using differential evolution to refine template parameters after initial upload, achieving a mean \SNR increase of $+1.64\%$ with a mean additional latency of 37 seconds; a scheme treating Virgo as a sky-map-only detector to improve sky localization while avoiding the sensitivity penalty of three-detector coincidence requirements; and an \EW search using frequency-truncated templates to issue pre-merger alerts for \BNS and \NSBH systems. The autogating procedure was also improved, extending glitch suppression from individual 8\,s analysis segments to the full rolling strain buffer, enabling more reliable removal of loud and closely spaced glitches.

Performance on the Mock Data Challenge demonstrated significant sensitivity gains over the O3 configuration. For the coincident search, sensitive volume improved by factors of 1.7--2.3 at an inverse false alarm rate of 10 years, with the largest gains at lower masses, and overall detection efficiency at a false alarm rate below one per year improved from 50.6\% to 79.3\%. For the single-detector search the improvement was smaller, with sensitive volume gains of 1.3--1.7 and efficiency improving from 14.5\% to 18.6\%. Upload latency requirements were met, with a median of 15.94 seconds from merger to candidate upload and 99.89\% of events uploaded within 50 seconds.

The \EW search was characterized through a dedicated injection study with 32,599 \BNS and \NSBH signals in simulated Gaussian noise. The study identified a timing accuracy limitation: at low frequency cutoffs, the effective bandwidth of truncated templates is small enough that the 1$\sigma$ timing error (up to 5.89 ms at 29 Hz) can exceed the 3 ms margin of the current 13 ms coincidence window, causing coincident trigger pairs to fall outside the window and preventing candidate formation. This affected 2.68\% of injections. The solutions (widening the coincidence window on a frequency-dependent basis, applying subsample interpolation, and retuning the phase-time-amplitude ranking histogram) have been identified and will be implemented in future observing runs, improving the sensitivity of the \EW search for multi-messenger events.

\begin{acknowledgments}
MT acknowledges support from the U.S. National Science Foundation through grant PHY-2409448. IWH and GCD acknowledge support from the STFC through grants ST/Y005260/1, ST/V005715/1 and UKRI2490. AT acknowledges support from the DISCNet doctoral training network. TD acknowledges support from Mar{'i}a de Maeztu grant CEX2023-001318-M funded by MICIU/AEI/10.13039/501100011033, and from Grant ED431F 2025/04 of the Galician Conselleria de Educacion, Ciencia, Universidades e Formacion Profesional.

The analyses described here made use of the PyCBC software library~\cite{pycbc-software}. This material is based upon work supported by NSF's LIGO Laboratory
which is a major facility fully funded by the National Science Foundation. The authors are grateful for computational resources provided by the LIGO Laboratory and supported by National Science Foundation Grants PHY-0757058 and PHY-0823459.

\end{acknowledgments}

\bibliography{references.bib}

\begin{thebibliography}{60}%
\makeatletter
\providecommand \@ifxundefined [1]{%
 \@ifx{#1\undefined}
}%
\providecommand \@ifnum [1]{%
 \ifnum #1\expandafter \@firstoftwo
 \else \expandafter \@secondoftwo
 \fi
}%
\providecommand \@ifx [1]{%
 \ifx #1\expandafter \@firstoftwo
 \else \expandafter \@secondoftwo
 \fi
}%
\providecommand \natexlab [1]{#1}%
\providecommand \enquote  [1]{``#1''}%
\providecommand \bibnamefont  [1]{#1}%
\providecommand \bibfnamefont [1]{#1}%
\providecommand \citenamefont [1]{#1}%
\providecommand \href@noop [0]{\@secondoftwo}%
\providecommand \href [0]{\begingroup \@sanitize@url \@href}%
\providecommand \@href[1]{\@@startlink{#1}\@@href}%
\providecommand \@@href[1]{\endgroup#1\@@endlink}%
\providecommand \@sanitize@url [0]{\catcode `\\12\catcode `\$12\catcode
  `\&12\catcode `\#12\catcode `\^12\catcode `\_12\catcode `\%12\relax}%
\providecommand \@@startlink[1]{}%
\providecommand \@@endlink[0]{}%
\providecommand \url  [0]{\begingroup\@sanitize@url \@url }%
\providecommand \@url [1]{\endgroup\@href {#1}{\urlprefix }}%
\providecommand \urlprefix  [0]{URL }%
\providecommand \Eprint [0]{\href }%
\providecommand \doibase [0]{https://doi.org/}%
\providecommand \selectlanguage [0]{\@gobble}%
\providecommand \bibinfo  [0]{\@secondoftwo}%
\providecommand \bibfield  [0]{\@secondoftwo}%
\providecommand \translation [1]{[#1]}%
\providecommand \BibitemOpen [0]{}%
\providecommand \bibitemStop [0]{}%
\providecommand \bibitemNoStop [0]{.\EOS\space}%
\providecommand \EOS [0]{\spacefactor3000\relax}%
\providecommand \BibitemShut  [1]{\csname bibitem#1\endcsname}%
\let\auto@bib@innerbib\@empty
%</preamble>
\bibitem [{\citenamefont {Abbott}\ \emph
  {et~al.}(2016{\natexlab{a}})\citenamefont {Abbott} \emph
  {et~al.}}]{LIGOScientific:2016vbw}%
  \BibitemOpen
  \bibfield  {author} {\bibinfo {author} {\bibfnamefont {B.~P.}\ \bibnamefont
  {Abbott}} \emph {et~al.} (\bibinfo {collaboration} {LIGO Scientific,
  Virgo}),\ }\bibfield  {title} {\bibinfo {title} {{GW150914: First results
  from the search for binary black hole coalescence with Advanced LIGO}},\
  }\href {https://doi.org/10.1103/PhysRevD.93.122003} {\bibfield  {journal}
  {\bibinfo  {journal} {Phys. Rev. D}\ }\textbf {\bibinfo {volume} {93}},\
  \bibinfo {pages} {122003} (\bibinfo {year} {2016}{\natexlab{a}})},\ \Eprint
  {https://arxiv.org/abs/1602.03839} {arXiv:1602.03839 [gr-qc]} \BibitemShut
  {NoStop}%
\bibitem [{\citenamefont {Aasi}\ \emph {et~al.}(2015)\citenamefont {Aasi} \emph
  {et~al.}}]{LIGOScientific:2014pky}%
  \BibitemOpen
  \bibfield  {author} {\bibinfo {author} {\bibfnamefont {J.}~\bibnamefont
  {Aasi}} \emph {et~al.} (\bibinfo {collaboration} {LIGO Scientific}),\
  }\bibfield  {title} {\bibinfo {title} {{Advanced LIGO}},\ }\href
  {https://doi.org/10.1088/0264-9381/32/7/074001} {\bibfield  {journal}
  {\bibinfo  {journal} {Class. Quant. Grav.}\ }\textbf {\bibinfo {volume}
  {32}},\ \bibinfo {pages} {074001} (\bibinfo {year} {2015})},\ \Eprint
  {https://arxiv.org/abs/1411.4547} {arXiv:1411.4547 [gr-qc]} \BibitemShut
  {NoStop}%
\bibitem [{\citenamefont {Acernese}\ \emph {et~al.}(2015)\citenamefont
  {Acernese} \emph {et~al.}}]{VIRGO:2014yos}%
  \BibitemOpen
  \bibfield  {author} {\bibinfo {author} {\bibfnamefont {F.}~\bibnamefont
  {Acernese}} \emph {et~al.} (\bibinfo {collaboration} {Virgo}),\ }\bibfield
  {title} {\bibinfo {title} {{Advanced Virgo: a second-generation
  interferometric gravitational wave detector}},\ }\href
  {https://doi.org/10.1088/0264-9381/32/2/024001} {\bibfield  {journal}
  {\bibinfo  {journal} {Class. Quant. Grav.}\ }\textbf {\bibinfo {volume}
  {32}},\ \bibinfo {pages} {024001} (\bibinfo {year} {2015})},\ \Eprint
  {https://arxiv.org/abs/1408.3978} {arXiv:1408.3978 [gr-qc]} \BibitemShut
  {NoStop}%
\bibitem [{\citenamefont {Akutsu}\ \emph {et~al.}(2021)\citenamefont {Akutsu}
  \emph {et~al.}}]{KAGRA:2020tym}%
  \BibitemOpen
  \bibfield  {author} {\bibinfo {author} {\bibfnamefont {T.}~\bibnamefont
  {Akutsu}} \emph {et~al.} (\bibinfo {collaboration} {KAGRA}),\ }\bibfield
  {title} {\bibinfo {title} {{Overview of KAGRA: Detector design and
  construction history}},\ }\href {https://doi.org/10.1093/ptep/ptaa125}
  {\bibfield  {journal} {\bibinfo  {journal} {PTEP}\ }\textbf {\bibinfo
  {volume} {2021}},\ \bibinfo {pages} {05A101} (\bibinfo {year} {2021})},\
  \Eprint {https://arxiv.org/abs/2005.05574} {arXiv:2005.05574
  [physics.ins-det]} \BibitemShut {NoStop}%
\bibitem [{\citenamefont {Somiya}(2012)}]{Somiya:2011np}%
  \BibitemOpen
  \bibfield  {author} {\bibinfo {author} {\bibfnamefont {K.}~\bibnamefont
  {Somiya}} (\bibinfo {collaboration} {KAGRA}),\ }\bibfield  {title} {\bibinfo
  {title} {{Detector configuration of KAGRA: The Japanese cryogenic
  gravitational-wave detector}},\ }\href
  {https://doi.org/10.1088/0264-9381/29/12/124007} {\bibfield  {journal}
  {\bibinfo  {journal} {Class. Quant. Grav.}\ }\textbf {\bibinfo {volume}
  {29}},\ \bibinfo {pages} {124007} (\bibinfo {year} {2012})},\ \Eprint
  {https://arxiv.org/abs/1111.7185} {arXiv:1111.7185 [gr-qc]} \BibitemShut
  {NoStop}%
\bibitem [{\citenamefont {Aso}\ \emph {et~al.}(2013)\citenamefont {Aso},
  \citenamefont {Michimura}, \citenamefont {Somiya}, \citenamefont {Ando},
  \citenamefont {Miyakawa}, \citenamefont {Sekiguchi}, \citenamefont
  {Tatsumi},\ and\ \citenamefont {Yamamoto}}]{Aso:2013eba}%
  \BibitemOpen
  \bibfield  {author} {\bibinfo {author} {\bibfnamefont {Y.}~\bibnamefont
  {Aso}}, \bibinfo {author} {\bibfnamefont {Y.}~\bibnamefont {Michimura}},
  \bibinfo {author} {\bibfnamefont {K.}~\bibnamefont {Somiya}}, \bibinfo
  {author} {\bibfnamefont {M.}~\bibnamefont {Ando}}, \bibinfo {author}
  {\bibfnamefont {O.}~\bibnamefont {Miyakawa}}, \bibinfo {author}
  {\bibfnamefont {T.}~\bibnamefont {Sekiguchi}}, \bibinfo {author}
  {\bibfnamefont {D.}~\bibnamefont {Tatsumi}},\ and\ \bibinfo {author}
  {\bibfnamefont {H.}~\bibnamefont {Yamamoto}} (\bibinfo {collaboration}
  {KAGRA}),\ }\bibfield  {title} {\bibinfo {title} {{Interferometer design of
  the KAGRA gravitational wave detector}},\ }\href
  {https://doi.org/10.1103/PhysRevD.88.043007} {\bibfield  {journal} {\bibinfo
  {journal} {Phys. Rev. D}\ }\textbf {\bibinfo {volume} {88}},\ \bibinfo
  {pages} {043007} (\bibinfo {year} {2013})},\ \Eprint
  {https://arxiv.org/abs/1306.6747} {arXiv:1306.6747 [gr-qc]} \BibitemShut
  {NoStop}%
\bibitem [{LIG(2026)}]{LIGOScientific:2026wfs}%
  \BibitemOpen
  \bibfield  {title} {\bibinfo {title} {{GWTC-5.0: Observations from the Second
  Part of the Fourth LIGO-Virgo-KAGRA Observing Run and Updates to the
  Gravitational-Wave Transient Catalog}},\ }\href@noop {} {\bibfield  {journal}
  {\bibinfo  {journal} {arXiv}\ } (\bibinfo {year} {2026})},\ \Eprint
  {https://arxiv.org/abs/2605.27225} {arXiv:2605.27225 [gr-qc]} \BibitemShut
  {NoStop}%
\bibitem [{\citenamefont {Abbott}\ \emph
  {et~al.}(2017{\natexlab{a}})\citenamefont {Abbott} \emph
  {et~al.}}]{GW170817MMA}%
  \BibitemOpen
  \bibfield  {author} {\bibinfo {author} {\bibfnamefont {B.~P.}\ \bibnamefont
  {Abbott}} \emph {et~al.} (\bibinfo {collaboration} {LIGO Scientific, Virgo,
  Fermi GBM, INTEGRAL, IceCube, AstroSat Cadmium Zinc Telluride Imager Team,
  IPN, Insight-Hxmt, ANTARES, Swift, AGILE Team, 1M2H Team, Dark Energy Camera
  GW-EM, DES, DLT40, GRAWITA, Fermi-LAT, ATCA, ASKAP, Las Cumbres Observatory
  Group, OzGrav, DWF (Deeper Wider Faster Program), AST3, CAASTRO, VINROUGE,
  MASTER, J-GEM, GROWTH, JAGWAR, CaltechNRAO, TTU-NRAO, NuSTAR, Pan-STARRS,
  MAXI Team, TZAC Consortium, KU, Nordic Optical Telescope, ePESSTO, GROND,
  Texas Tech University, SALT Group, TOROS, BOOTES, MWA, CALET, IKI-GW
  Follow-up, H.E.S.S., LOFAR, LWA, HAWC, Pierre Auger, ALMA, Euro VLBI Team, Pi
  of Sky, Chandra Team at McGill University, DFN, ATLAS Telescopes, High Time
  Resolution Universe Survey, RIMAS, RATIR, SKA South Africa/MeerKAT}),\
  }\bibfield  {title} {\bibinfo {title} {{Multi-messenger Observations of a
  Binary Neutron Star Merger}},\ }\href
  {https://doi.org/10.3847/2041-8213/aa91c9} {\bibfield  {journal} {\bibinfo
  {journal} {Astrophys. J. Lett.}\ }\textbf {\bibinfo {volume} {848}},\
  \bibinfo {pages} {L12} (\bibinfo {year} {2017}{\natexlab{a}})},\ \Eprint
  {https://arxiv.org/abs/1710.05833} {arXiv:1710.05833 [astro-ph.HE]}
  \BibitemShut {NoStop}%
\bibitem [{\citenamefont {Mishra}\ \emph {et~al.}(2025)\citenamefont {Mishra},
  \citenamefont {Bhaumik}, \citenamefont {Gayathri}, \citenamefont
  {Szczepa{\'n}czyk}, \citenamefont {Bartos},\ and\ \citenamefont
  {Klimenko}}]{Mishra:2024zzs}%
  \BibitemOpen
  \bibfield  {author} {\bibinfo {author} {\bibfnamefont {T.}~\bibnamefont
  {Mishra}}, \bibinfo {author} {\bibfnamefont {S.}~\bibnamefont {Bhaumik}},
  \bibinfo {author} {\bibfnamefont {V.}~\bibnamefont {Gayathri}}, \bibinfo
  {author} {\bibfnamefont {M.~J.}\ \bibnamefont {Szczepa{\'n}czyk}}, \bibinfo
  {author} {\bibfnamefont {I.}~\bibnamefont {Bartos}},\ and\ \bibinfo {author}
  {\bibfnamefont {S.}~\bibnamefont {Klimenko}},\ }\bibfield  {title} {\bibinfo
  {title} {{Gravitational waves detected by a burst search in
  LIGO/Virgo{\textquoteright}s third observing run}},\ }\href
  {https://doi.org/10.1103/PhysRevD.111.023054} {\bibfield  {journal} {\bibinfo
   {journal} {Phys. Rev. D}\ }\textbf {\bibinfo {volume} {111}},\ \bibinfo
  {pages} {023054} (\bibinfo {year} {2025})},\ \Eprint
  {https://arxiv.org/abs/2410.15191} {arXiv:2410.15191 [astro-ph.HE]}
  \BibitemShut {NoStop}%
\bibitem [{\citenamefont {Ewing}\ \emph {et~al.}(2024)\citenamefont {Ewing}
  \emph {et~al.}}]{Ewing:2023qqe}%
  \BibitemOpen
  \bibfield  {author} {\bibinfo {author} {\bibfnamefont {B.}~\bibnamefont
  {Ewing}} \emph {et~al.},\ }\bibfield  {title} {\bibinfo {title} {{Performance
  of the low-latency GstLAL inspiral search towards LIGO, Virgo, and
  KAGRA{\textquoteright}s fourth observing run}},\ }\href
  {https://doi.org/10.1103/PhysRevD.109.042008} {\bibfield  {journal} {\bibinfo
   {journal} {Phys. Rev. D}\ }\textbf {\bibinfo {volume} {109}},\ \bibinfo
  {pages} {042008} (\bibinfo {year} {2024})},\ \Eprint
  {https://arxiv.org/abs/2305.05625} {arXiv:2305.05625 [gr-qc]} \BibitemShut
  {NoStop}%
\bibitem [{\citenamefont {All{\'e}n{\'e}}\ \emph {et~al.}(2025)\citenamefont
  {All{\'e}n{\'e}} \emph {et~al.}}]{Allene:2025saz}%
  \BibitemOpen
  \bibfield  {author} {\bibinfo {author} {\bibfnamefont {C.}~\bibnamefont
  {All{\'e}n{\'e}}} \emph {et~al.},\ }\bibfield  {title} {\bibinfo {title}
  {{The MBTA pipeline for detecting compact binary coalescences in the fourth
  LIGO-Virgo-KAGRA observing run}},\ }\href
  {https://doi.org/10.1088/1361-6382/add234} {\bibfield  {journal} {\bibinfo
  {journal} {Class. Quant. Grav.}\ }\textbf {\bibinfo {volume} {42}},\ \bibinfo
  {pages} {105009} (\bibinfo {year} {2025})},\ \Eprint
  {https://arxiv.org/abs/2501.04598} {arXiv:2501.04598 [gr-qc]} \BibitemShut
  {NoStop}%
\bibitem [{\citenamefont {Nitz}\ \emph {et~al.}(2018)\citenamefont {Nitz},
  \citenamefont {Dal~Canton}, \citenamefont {Davis},\ and\ \citenamefont
  {Reyes}}]{PyCBCLiveO2}%
  \BibitemOpen
  \bibfield  {author} {\bibinfo {author} {\bibfnamefont {A.~H.}\ \bibnamefont
  {Nitz}}, \bibinfo {author} {\bibfnamefont {T.}~\bibnamefont {Dal~Canton}},
  \bibinfo {author} {\bibfnamefont {D.}~\bibnamefont {Davis}},\ and\ \bibinfo
  {author} {\bibfnamefont {S.}~\bibnamefont {Reyes}},\ }\bibfield  {title}
  {\bibinfo {title} {{Rapid detection of gravitational waves from compact
  binary mergers with PyCBC Live}},\ }\href
  {https://doi.org/10.1103/PhysRevD.98.024050} {\bibfield  {journal} {\bibinfo
  {journal} {\prd}\ }\textbf {\bibinfo {volume} {98}},\ \bibinfo {pages}
  {024050} (\bibinfo {year} {2018})},\ \Eprint
  {https://arxiv.org/abs/1805.11174} {arXiv:1805.11174 [gr-qc]} \BibitemShut
  {NoStop}%
%%CITATION = ARXIV:1805.11174;%%
\bibitem [{\citenamefont {Dal~Canton}\ \emph {et~al.}(2021)\citenamefont
  {Dal~Canton}, \citenamefont {Nitz}, \citenamefont {Gadre}, \citenamefont
  {Cabourn~Davies}, \citenamefont {Villa-Ortega}, \citenamefont {Dent},
  \citenamefont {Harry},\ and\ \citenamefont {Xiao}}]{PyCBCLiveO3}%
  \BibitemOpen
  \bibfield  {author} {\bibinfo {author} {\bibfnamefont {T.}~\bibnamefont
  {Dal~Canton}}, \bibinfo {author} {\bibfnamefont {A.~H.}\ \bibnamefont
  {Nitz}}, \bibinfo {author} {\bibfnamefont {B.}~\bibnamefont {Gadre}},
  \bibinfo {author} {\bibfnamefont {G.~S.}\ \bibnamefont {Cabourn~Davies}},
  \bibinfo {author} {\bibfnamefont {V.}~\bibnamefont {Villa-Ortega}}, \bibinfo
  {author} {\bibfnamefont {T.}~\bibnamefont {Dent}}, \bibinfo {author}
  {\bibfnamefont {I.}~\bibnamefont {Harry}},\ and\ \bibinfo {author}
  {\bibfnamefont {L.}~\bibnamefont {Xiao}},\ }\bibfield  {title} {\bibinfo
  {title} {{Real-time Search for Compact Binary Mergers in Advanced LIGO and
  Virgo's Third Observing Run Using PyCBC Live}},\ }\href
  {https://doi.org/10.3847/1538-4357/ac2f9a} {\bibfield  {journal} {\bibinfo
  {journal} {Astrophys. J.}\ }\textbf {\bibinfo {volume} {923}},\ \bibinfo
  {pages} {254} (\bibinfo {year} {2021})},\ \Eprint
  {https://arxiv.org/abs/2008.07494} {arXiv:2008.07494 [astro-ph.HE]}
  \BibitemShut {NoStop}%
\bibitem [{\citenamefont {Chu}\ \emph {et~al.}(2022)\citenamefont {Chu} \emph
  {et~al.}}]{Chu:2020pjv}%
  \BibitemOpen
  \bibfield  {author} {\bibinfo {author} {\bibfnamefont {Q.}~\bibnamefont
  {Chu}} \emph {et~al.},\ }\bibfield  {title} {\bibinfo {title} {{SPIIR online
  coherent pipeline to search for gravitational waves from compact binary
  coalescences}},\ }\href {https://doi.org/10.1103/PhysRevD.105.024023}
  {\bibfield  {journal} {\bibinfo  {journal} {Phys. Rev. D}\ }\textbf {\bibinfo
  {volume} {105}},\ \bibinfo {pages} {024023} (\bibinfo {year} {2022})},\
  \Eprint {https://arxiv.org/abs/2011.06787} {arXiv:2011.06787 [gr-qc]}
  \BibitemShut {NoStop}%
\bibitem [{\citenamefont {Singer}\ and\ \citenamefont
  {Price}(2016)}]{Singer:2015ema}%
  \BibitemOpen
  \bibfield  {author} {\bibinfo {author} {\bibfnamefont {L.~P.}\ \bibnamefont
  {Singer}}\ and\ \bibinfo {author} {\bibfnamefont {L.~R.}\ \bibnamefont
  {Price}},\ }\bibfield  {title} {\bibinfo {title} {{Rapid Bayesian position
  reconstruction for gravitational-wave transients}},\ }\href
  {https://doi.org/10.1103/PhysRevD.93.024013} {\bibfield  {journal} {\bibinfo
  {journal} {Phys. Rev. D}\ }\textbf {\bibinfo {volume} {93}},\ \bibinfo
  {pages} {024013} (\bibinfo {year} {2016})},\ \Eprint
  {https://arxiv.org/abs/1508.03634} {arXiv:1508.03634 [gr-qc]} \BibitemShut
  {NoStop}%
\bibitem [{\citenamefont {Capote}\ \emph {et~al.}(2025)\citenamefont {Capote}
  \emph {et~al.}}]{Capote:2024rmo}%
  \BibitemOpen
  \bibfield  {author} {\bibinfo {author} {\bibfnamefont {E.}~\bibnamefont
  {Capote}} \emph {et~al.},\ }\bibfield  {title} {\bibinfo {title} {{Advanced
  LIGO detector performance in the fourth observing run}},\ }\href
  {https://doi.org/10.1103/PhysRevD.111.062002} {\bibfield  {journal} {\bibinfo
   {journal} {Phys. Rev. D}\ }\textbf {\bibinfo {volume} {111}},\ \bibinfo
  {pages} {062002} (\bibinfo {year} {2025})},\ \Eprint
  {https://arxiv.org/abs/2411.14607} {arXiv:2411.14607 [gr-qc]} \BibitemShut
  {NoStop}%
\bibitem [{\citenamefont {Acernese}\ \emph {et~al.}(2019)\citenamefont
  {Acernese} \emph {et~al.}}]{Virgo:2019juy}%
  \BibitemOpen
  \bibfield  {author} {\bibinfo {author} {\bibfnamefont {F.}~\bibnamefont
  {Acernese}} \emph {et~al.} (\bibinfo {collaboration} {Virgo}),\ }\bibfield
  {title} {\bibinfo {title} {{Increasing the Astrophysical Reach of the
  Advanced Virgo Detector via the Application of Squeezed Vacuum States of
  Light}},\ }\href {https://doi.org/10.1103/PhysRevLett.123.231108} {\bibfield
  {journal} {\bibinfo  {journal} {Phys. Rev. Lett.}\ }\textbf {\bibinfo
  {volume} {123}},\ \bibinfo {pages} {231108} (\bibinfo {year}
  {2019})}\BibitemShut {NoStop}%
\bibitem [{\citenamefont {Abac}\ \emph
  {et~al.}(2025{\natexlab{a}})\citenamefont {Abac} \emph
  {et~al.}}]{LIGOScientific:2025slb}%
  \BibitemOpen
  \bibfield  {author} {\bibinfo {author} {\bibfnamefont {A.~G.}\ \bibnamefont
  {Abac}} \emph {et~al.} (\bibinfo {collaboration} {LIGO Scientific, VIRGO,
  KAGRA}),\ }\bibfield  {title} {\bibinfo {title} {{GWTC-4.0: Updating the
  Gravitational-Wave Transient Catalog with Observations from the First Part of
  the Fourth LIGO-Virgo-KAGRA Observing Run}},\ }\href@noop {} {\bibfield
  {journal} {\bibinfo  {journal} {Astrophys. J. Lett.}\ } (\bibinfo {year}
  {2025}{\natexlab{a}})},\ \Eprint {https://arxiv.org/abs/2508.18082}
  {arXiv:2508.18082 [gr-qc]} \BibitemShut {NoStop}%
\bibitem [{\citenamefont {Abac}\ \emph
  {et~al.}(2025{\natexlab{b}})\citenamefont {Abac} \emph
  {et~al.}}]{LIGOScientific:2025hdt}%
  \BibitemOpen
  \bibfield  {author} {\bibinfo {author} {\bibfnamefont {A.~G.}\ \bibnamefont
  {Abac}} \emph {et~al.} (\bibinfo {collaboration} {LIGO Scientific, KAGRA,
  VIRGO}),\ }\bibfield  {title} {\bibinfo {title} {{GWTC-4.0: An Introduction
  to Version 4.0 of the Gravitational-Wave Transient Catalog}},\ }\href
  {https://doi.org/10.3847/2041-8213/ae0c06} {\bibfield  {journal} {\bibinfo
  {journal} {Astrophys. J. Lett.}\ }\textbf {\bibinfo {volume} {995}},\
  \bibinfo {pages} {L18} (\bibinfo {year} {2025}{\natexlab{b}})},\ \Eprint
  {https://arxiv.org/abs/2508.18080} {arXiv:2508.18080 [gr-qc]} \BibitemShut
  {NoStop}%
\bibitem [{\citenamefont {Abac}\ \emph {et~al.}(2024)\citenamefont {Abac} \emph
  {et~al.}}]{LIGOScientific:2024elc}%
  \BibitemOpen
  \bibfield  {author} {\bibinfo {author} {\bibfnamefont {A.~G.}\ \bibnamefont
  {Abac}} \emph {et~al.} (\bibinfo {collaboration} {LIGO Scientific, KAGRA,
  VIRGO}),\ }\bibfield  {title} {\bibinfo {title} {{Observation of
  Gravitational Waves from the Coalescence of a 2.5{\textendash}4.5 M
  $_{\odot}$ Compact Object and a Neutron Star}},\ }\href
  {https://doi.org/10.3847/2041-8213/ad5beb} {\bibfield  {journal} {\bibinfo
  {journal} {Astrophys. J. Lett.}\ }\textbf {\bibinfo {volume} {970}},\
  \bibinfo {pages} {L34} (\bibinfo {year} {2024})},\ \Eprint
  {https://arxiv.org/abs/2404.04248} {arXiv:2404.04248 [astro-ph.HE]}
  \BibitemShut {NoStop}%
\bibitem [{\citenamefont {Abac}\ \emph
  {et~al.}(2025{\natexlab{c}})\citenamefont {Abac} \emph
  {et~al.}}]{LIGOScientific:2025rsn}%
  \BibitemOpen
  \bibfield  {author} {\bibinfo {author} {\bibfnamefont {A.~G.}\ \bibnamefont
  {Abac}} \emph {et~al.} (\bibinfo {collaboration} {LIGO Scientific, VIRGO,
  KAGRA}),\ }\bibfield  {title} {\bibinfo {title} {{GW231123: A Binary Black
  Hole Merger with Total Mass 190{\textendash}265 M$_{\odot}$}},\ }\href
  {https://doi.org/10.3847/2041-8213/ae0c9c} {\bibfield  {journal} {\bibinfo
  {journal} {Astrophys. J. Lett.}\ }\textbf {\bibinfo {volume} {993}},\
  \bibinfo {pages} {L25} (\bibinfo {year} {2025}{\natexlab{c}})},\ \Eprint
  {https://arxiv.org/abs/2507.08219} {arXiv:2507.08219 [astro-ph.HE]}
  \BibitemShut {NoStop}%
\bibitem [{\citenamefont {Abac}\ \emph
  {et~al.}(2025{\natexlab{d}})\citenamefont {Abac} \emph
  {et~al.}}]{LIGOScientific:2025brd}%
  \BibitemOpen
  \bibfield  {author} {\bibinfo {author} {\bibfnamefont {A.~G.}\ \bibnamefont
  {Abac}} \emph {et~al.} (\bibinfo {collaboration} {LIGO Scientific, Virgo,
  KAGRA}),\ }\bibfield  {title} {\bibinfo {title} {{GW241011 and GW241110:
  Exploring Binary Formation and Fundamental Physics with Asymmetric, High-spin
  Black Hole Coalescences}},\ }\href {https://doi.org/10.3847/2041-8213/ae0d54}
  {\bibfield  {journal} {\bibinfo  {journal} {Astrophys. J. Lett.}\ }\textbf
  {\bibinfo {volume} {993}},\ \bibinfo {pages} {L21} (\bibinfo {year}
  {2025}{\natexlab{d}})},\ \Eprint {https://arxiv.org/abs/2510.26931}
  {arXiv:2510.26931 [astro-ph.HE]} \BibitemShut {NoStop}%
\bibitem [{\citenamefont {Abac}\ \emph
  {et~al.}(2025{\natexlab{e}})\citenamefont {Abac} \emph
  {et~al.}}]{LIGOScientific:2025rid}%
  \BibitemOpen
  \bibfield  {author} {\bibinfo {author} {\bibfnamefont {A.~G.}\ \bibnamefont
  {Abac}} \emph {et~al.} (\bibinfo {collaboration} {LIGO Scientific, Virgo,
  KAGRA}),\ }\bibfield  {title} {\bibinfo {title} {{GW250114: Testing
  Hawking{\textquoteright}s Area Law and the Kerr Nature of Black Holes}},\
  }\href {https://doi.org/10.1103/kw5g-d732} {\bibfield  {journal} {\bibinfo
  {journal} {Phys. Rev. Lett.}\ }\textbf {\bibinfo {volume} {135}},\ \bibinfo
  {pages} {111403} (\bibinfo {year} {2025}{\natexlab{e}})},\ \Eprint
  {https://arxiv.org/abs/2509.08054} {arXiv:2509.08054 [gr-qc]} \BibitemShut
  {NoStop}%
\bibitem [{\citenamefont {Chaudhary}\ \emph {et~al.}(2023)\citenamefont
  {Chaudhary} \emph {et~al.}}]{Chaudhary:2023vec}%
  \BibitemOpen
  \bibfield  {author} {\bibinfo {author} {\bibfnamefont {S.~S.}\ \bibnamefont
  {Chaudhary}} \emph {et~al.},\ }\bibfield  {title} {\bibinfo {title}
  {{Low-latency gravitational wave alert products and their performance in
  anticipation of the fourth LIGO-Virgo-KAGRA observing run}},\ }\href@noop {}
  {\bibfield  {journal} {\bibinfo  {journal} {arxiv e-prints}\ } (\bibinfo
  {year} {2023})},\ \Eprint {https://arxiv.org/abs/2308.04545}
  {arXiv:2308.04545 [astro-ph.HE]} \BibitemShut {NoStop}%
\bibitem [{\citenamefont {Nitz}\ \emph {et~al.}(2020)\citenamefont {Nitz},
  \citenamefont {Sch{\"a}fer},\ and\ \citenamefont
  {Dal~Canton}}]{Nitz:2020vym}%
  \BibitemOpen
  \bibfield  {author} {\bibinfo {author} {\bibfnamefont {A.~H.}\ \bibnamefont
  {Nitz}}, \bibinfo {author} {\bibfnamefont {M.}~\bibnamefont {Sch{\"a}fer}},\
  and\ \bibinfo {author} {\bibfnamefont {T.}~\bibnamefont {Dal~Canton}},\
  }\bibfield  {title} {\bibinfo {title} {{Gravitational-wave Merger
  Forecasting: Scenarios for the Early Detection and Localization of
  Compact-binary Mergers with Ground-based Observatories}},\ }\href
  {https://doi.org/10.3847/2041-8213/abbc10} {\bibfield  {journal} {\bibinfo
  {journal} {Astrophys. J. Lett.}\ }\textbf {\bibinfo {volume} {902}},\
  \bibinfo {pages} {L29} (\bibinfo {year} {2020})},\ \Eprint
  {https://arxiv.org/abs/2009.04439} {arXiv:2009.04439 [astro-ph.HE]}
  \BibitemShut {NoStop}%
\bibitem [{\citenamefont {Magee}\ \emph {et~al.}(2021)\citenamefont {Magee}
  \emph {et~al.}}]{Magee:2021xdx}%
  \BibitemOpen
  \bibfield  {author} {\bibinfo {author} {\bibfnamefont {R.}~\bibnamefont
  {Magee}} \emph {et~al.},\ }\bibfield  {title} {\bibinfo {title} {{First
  demonstration of early warning gravitational wave alerts}},\ }\href
  {https://doi.org/10.3847/2041-8213/abed54} {\bibfield  {journal} {\bibinfo
  {journal} {Astrophys. J. Lett.}\ }\textbf {\bibinfo {volume} {910}},\
  \bibinfo {pages} {L21} (\bibinfo {year} {2021})},\ \Eprint
  {https://arxiv.org/abs/2102.04555} {arXiv:2102.04555 [astro-ph.HE]}
  \BibitemShut {NoStop}%
\bibitem [{\citenamefont {Soni}\ \emph {et~al.}(2025)\citenamefont {Soni} \emph
  {et~al.}}]{LIGO:2024kkz}%
  \BibitemOpen
  \bibfield  {author} {\bibinfo {author} {\bibfnamefont {S.}~\bibnamefont
  {Soni}} \emph {et~al.} (\bibinfo {collaboration} {LIGO}),\ }\bibfield
  {title} {\bibinfo {title} {{LIGO Detector Characterization in the first half
  of the fourth Observing run}},\ }\href
  {https://doi.org/10.1088/1361-6382/adc4b6} {\bibfield  {journal} {\bibinfo
  {journal} {Class. Quant. Grav.}\ }\textbf {\bibinfo {volume} {42}},\ \bibinfo
  {pages} {085016} (\bibinfo {year} {2025})},\ \Eprint
  {https://arxiv.org/abs/2409.02831} {arXiv:2409.02831 [astro-ph.IM]}
  \BibitemShut {NoStop}%
\bibitem [{\citenamefont {Yarbrough}\ \emph {et~al.}(2025)\citenamefont
  {Yarbrough}, \citenamefont {Guimaraes}, \citenamefont {Joshi}, \citenamefont
  {Gonz{\'a}lez}, \citenamefont {Valentini},\ and\ \citenamefont
  {Shah}}]{Yarbrough:2025nsa}%
  \BibitemOpen
  \bibfield  {author} {\bibinfo {author} {\bibfnamefont {Z.}~\bibnamefont
  {Yarbrough}}, \bibinfo {author} {\bibfnamefont {A.}~\bibnamefont
  {Guimaraes}}, \bibinfo {author} {\bibfnamefont {P.}~\bibnamefont {Joshi}},
  \bibinfo {author} {\bibfnamefont {G.}~\bibnamefont {Gonz{\'a}lez}}, \bibinfo
  {author} {\bibfnamefont {A.}~\bibnamefont {Valentini}},\ and\ \bibinfo
  {author} {\bibfnamefont {U.}~\bibnamefont {Shah}},\ }\bibfield  {title}
  {\bibinfo {title} {{PINCH: pipeline-informed noise characterization in
  LIGO{\textquoteright}s third observing run}},\ }\href
  {https://doi.org/10.1088/1361-6382/adf58c} {\bibfield  {journal} {\bibinfo
  {journal} {Class. Quant. Grav.}\ }\textbf {\bibinfo {volume} {42}},\ \bibinfo
  {pages} {165014} (\bibinfo {year} {2025})},\ \Eprint
  {https://arxiv.org/abs/2505.14949} {arXiv:2505.14949 [gr-qc]} \BibitemShut
  {NoStop}%
\bibitem [{\citenamefont {Essick}\ \emph {et~al.}(2020)\citenamefont {Essick},
  \citenamefont {Godwin}, \citenamefont {Hanna}, \citenamefont {Blackburn},\
  and\ \citenamefont {Katsavounidis}}]{Essick:2020qpo}%
  \BibitemOpen
  \bibfield  {author} {\bibinfo {author} {\bibfnamefont {R.}~\bibnamefont
  {Essick}}, \bibinfo {author} {\bibfnamefont {P.}~\bibnamefont {Godwin}},
  \bibinfo {author} {\bibfnamefont {C.}~\bibnamefont {Hanna}}, \bibinfo
  {author} {\bibfnamefont {L.}~\bibnamefont {Blackburn}},\ and\ \bibinfo
  {author} {\bibfnamefont {E.}~\bibnamefont {Katsavounidis}},\ }\bibfield
  {title} {\bibinfo {title} {{iDQ: Statistical Inference of Non-Gaussian Noise
  with Auxiliary Degrees of Freedom in Gravitational-Wave Detectors}},\ }\href
  {https://doi.org/10.1088/2632-2153/abab5f} {\bibfield  {journal} {\bibinfo
  {journal} {Mach. Learn. Sci. Technol.}\ }\textbf {\bibinfo {volume} {2}},\
  \bibinfo {pages} {015004} (\bibinfo {year} {2020})},\ \Eprint
  {https://arxiv.org/abs/2005.12761} {arXiv:2005.12761 [astro-ph.IM]}
  \BibitemShut {NoStop}%
\bibitem [{\citenamefont {Allen}(2005)}]{Allen:2004gu}%
  \BibitemOpen
  \bibfield  {author} {\bibinfo {author} {\bibfnamefont {B.}~\bibnamefont
  {Allen}},\ }\bibfield  {title} {\bibinfo {title} {{${\chi}^{2}$
  time-frequency discriminator for gravitational wave detection}},\ }\href
  {https://doi.org/10.1103/PhysRevD.71.062001} {\bibfield  {journal} {\bibinfo
  {journal} {Phys. Rev. D}\ }\textbf {\bibinfo {volume} {71}},\ \bibinfo
  {pages} {062001} (\bibinfo {year} {2005})},\ \Eprint
  {https://arxiv.org/abs/gr-qc/0405045} {arXiv:gr-qc/0405045} \BibitemShut
  {NoStop}%
\bibitem [{\citenamefont {Nitz}(2018)}]{Nitz:2017lco}%
  \BibitemOpen
  \bibfield  {author} {\bibinfo {author} {\bibfnamefont {A.~H.}\ \bibnamefont
  {Nitz}},\ }\bibfield  {title} {\bibinfo {title} {{Distinguishing short
  duration noise transients in LIGO data to improve the PyCBC search for
  gravitational waves from high mass binary black hole mergers}},\ }\href
  {https://doi.org/10.1088/1361-6382/aaa13d} {\bibfield  {journal} {\bibinfo
  {journal} {Class. Quant. Grav.}\ }\textbf {\bibinfo {volume} {35}},\ \bibinfo
  {pages} {035016} (\bibinfo {year} {2018})},\ \Eprint
  {https://arxiv.org/abs/1709.08974} {arXiv:1709.08974 [gr-qc]} \BibitemShut
  {NoStop}%
\bibitem [{\citenamefont {Davis}\ \emph {et~al.}(2022)\citenamefont {Davis},
  \citenamefont {Trevor}, \citenamefont {Mozzon},\ and\ \citenamefont
  {Nuttall}}]{Davis:2022cmw}%
  \BibitemOpen
  \bibfield  {author} {\bibinfo {author} {\bibfnamefont {D.}~\bibnamefont
  {Davis}}, \bibinfo {author} {\bibfnamefont {M.}~\bibnamefont {Trevor}},
  \bibinfo {author} {\bibfnamefont {S.}~\bibnamefont {Mozzon}},\ and\ \bibinfo
  {author} {\bibfnamefont {L.~K.}\ \bibnamefont {Nuttall}},\ }\bibfield
  {title} {\bibinfo {title} {{Incorporating information from LIGO data quality
  streams into the PyCBC search for gravitational waves}},\ }\href
  {https://doi.org/10.1103/PhysRevD.106.102006} {\bibfield  {journal} {\bibinfo
   {journal} {Phys. Rev. D}\ }\textbf {\bibinfo {volume} {106}},\ \bibinfo
  {pages} {102006} (\bibinfo {year} {2022})},\ \Eprint
  {https://arxiv.org/abs/2204.03091} {arXiv:2204.03091 [gr-qc]} \BibitemShut
  {NoStop}%
\bibitem [{\citenamefont {Davies}\ \emph {et~al.}(2020)\citenamefont {Davies},
  \citenamefont {Dent}, \citenamefont {T\'apai}, \citenamefont {Harry},
  \citenamefont {McIsaac},\ and\ \citenamefont {Nitz}}]{Davies:2020tsx}%
  \BibitemOpen
  \bibfield  {author} {\bibinfo {author} {\bibfnamefont {G.~S.}\ \bibnamefont
  {Davies}}, \bibinfo {author} {\bibfnamefont {T.}~\bibnamefont {Dent}},
  \bibinfo {author} {\bibfnamefont {M.}~\bibnamefont {T\'apai}}, \bibinfo
  {author} {\bibfnamefont {I.}~\bibnamefont {Harry}}, \bibinfo {author}
  {\bibfnamefont {C.}~\bibnamefont {McIsaac}},\ and\ \bibinfo {author}
  {\bibfnamefont {A.~H.}\ \bibnamefont {Nitz}},\ }\bibfield  {title} {\bibinfo
  {title} {{Extending the PyCBC search for gravitational waves from compact
  binary mergers to a global network}},\ }\href
  {https://doi.org/10.1103/PhysRevD.102.022004} {\bibfield  {journal} {\bibinfo
   {journal} {Phys. Rev. D}\ }\textbf {\bibinfo {volume} {102}},\ \bibinfo
  {pages} {022004} (\bibinfo {year} {2020})},\ \Eprint
  {https://arxiv.org/abs/2002.08291} {arXiv:2002.08291 [astro-ph.HE]}
  \BibitemShut {NoStop}%
\bibitem [{\citenamefont {Nitz}\ \emph {et~al.}(2017)\citenamefont {Nitz},
  \citenamefont {Dent}, \citenamefont {Dal~Canton}, \citenamefont {Fairhurst},\
  and\ \citenamefont {Brown}}]{Nitz:2017svb}%
  \BibitemOpen
  \bibfield  {author} {\bibinfo {author} {\bibfnamefont {A.~H.}\ \bibnamefont
  {Nitz}}, \bibinfo {author} {\bibfnamefont {T.}~\bibnamefont {Dent}}, \bibinfo
  {author} {\bibfnamefont {T.}~\bibnamefont {Dal~Canton}}, \bibinfo {author}
  {\bibfnamefont {S.}~\bibnamefont {Fairhurst}},\ and\ \bibinfo {author}
  {\bibfnamefont {D.~A.}\ \bibnamefont {Brown}},\ }\bibfield  {title} {\bibinfo
  {title} {{Detecting binary compact-object mergers with gravitational waves:
  Understanding and Improving the sensitivity of the PyCBC search}},\ }\href
  {https://doi.org/10.3847/1538-4357/aa8f50} {\bibfield  {journal} {\bibinfo
  {journal} {Astrophys. J.}\ }\textbf {\bibinfo {volume} {849}},\ \bibinfo
  {pages} {118} (\bibinfo {year} {2017})},\ \Eprint
  {https://arxiv.org/abs/1705.01513} {arXiv:1705.01513 [gr-qc]} \BibitemShut
  {NoStop}%
\bibitem [{\citenamefont {Usman}\ \emph {et~al.}(2016)\citenamefont {Usman}
  \emph {et~al.}}]{Usman:2015kfa}%
  \BibitemOpen
  \bibfield  {author} {\bibinfo {author} {\bibfnamefont {S.~A.}\ \bibnamefont
  {Usman}} \emph {et~al.},\ }\bibfield  {title} {\bibinfo {title} {{The PyCBC
  search for gravitational waves from compact binary coalescence}},\ }\href
  {https://doi.org/10.1088/0264-9381/33/21/215004} {\bibfield  {journal}
  {\bibinfo  {journal} {Class. Quant. Grav.}\ }\textbf {\bibinfo {volume}
  {33}},\ \bibinfo {pages} {215004} (\bibinfo {year} {2016})},\ \Eprint
  {https://arxiv.org/abs/1508.02357} {arXiv:1508.02357 [gr-qc]} \BibitemShut
  {NoStop}%
\bibitem [{\citenamefont {Cabourn~Davies}\ and\ \citenamefont
  {Harry}(2022)}]{Davies:2022thw}%
  \BibitemOpen
  \bibfield  {author} {\bibinfo {author} {\bibfnamefont {G.~S.}\ \bibnamefont
  {Cabourn~Davies}}\ and\ \bibinfo {author} {\bibfnamefont {I.~W.}\
  \bibnamefont {Harry}},\ }\bibfield  {title} {\bibinfo {title} {{Establishing
  significance of gravitational-wave signals from a single observatory in the
  PyCBC offline search}},\ }\href {https://doi.org/10.1088/1361-6382/ac8862}
  {\bibfield  {journal} {\bibinfo  {journal} {Class. Quant. Grav.}\ }\textbf
  {\bibinfo {volume} {39}},\ \bibinfo {pages} {215012} (\bibinfo {year}
  {2022})},\ \Eprint {https://arxiv.org/abs/2203.08545} {arXiv:2203.08545
  [gr-qc]} \BibitemShut {NoStop}%
\bibitem [{\citenamefont {Abbott}\ \emph
  {et~al.}(2016{\natexlab{b}})\citenamefont {Abbott} \emph
  {et~al.}}]{LIGOScientific:2016kwr}%
  \BibitemOpen
  \bibfield  {author} {\bibinfo {author} {\bibfnamefont {B.~P.}\ \bibnamefont
  {Abbott}} \emph {et~al.} (\bibinfo {collaboration} {LIGO Scientific,
  Virgo}),\ }\bibfield  {title} {\bibinfo {title} {{The Rate of Binary Black
  Hole Mergers Inferred from Advanced LIGO Observations Surrounding
  GW150914}},\ }\href {https://doi.org/10.3847/2041-8205/833/1/L1} {\bibfield
  {journal} {\bibinfo  {journal} {Astrophys. J. Lett.}\ }\textbf {\bibinfo
  {volume} {833}},\ \bibinfo {pages} {L1} (\bibinfo {year}
  {2016}{\natexlab{b}})},\ \Eprint {https://arxiv.org/abs/1602.03842}
  {arXiv:1602.03842 [astro-ph.HE]} \BibitemShut {NoStop}%
\bibitem [{\citenamefont {Lynch}\ \emph {et~al.}(2018)\citenamefont {Lynch},
  \citenamefont {Coughlin}, \citenamefont {Vitale}, \citenamefont {Stubbs},\
  and\ \citenamefont {Katsavounidis}}]{Lynch:2018yom}%
  \BibitemOpen
  \bibfield  {author} {\bibinfo {author} {\bibfnamefont {R.}~\bibnamefont
  {Lynch}}, \bibinfo {author} {\bibfnamefont {M.}~\bibnamefont {Coughlin}},
  \bibinfo {author} {\bibfnamefont {S.}~\bibnamefont {Vitale}}, \bibinfo
  {author} {\bibfnamefont {C.~W.}\ \bibnamefont {Stubbs}},\ and\ \bibinfo
  {author} {\bibfnamefont {E.}~\bibnamefont {Katsavounidis}},\ }\bibfield
  {title} {\bibinfo {title} {{Observational implications of lowering the
  LIGO-Virgo alert threshold}},\ }\href
  {https://doi.org/10.3847/2041-8213/aacf9f} {\bibfield  {journal} {\bibinfo
  {journal} {Astrophys. J. Lett.}\ }\textbf {\bibinfo {volume} {861}},\
  \bibinfo {pages} {L24} (\bibinfo {year} {2018})},\ \Eprint
  {https://arxiv.org/abs/1803.02880} {arXiv:1803.02880 [astro-ph.HE]}
  \BibitemShut {NoStop}%
\bibitem [{\citenamefont {Schutz}(2011)}]{Schutz:2011tw}%
  \BibitemOpen
  \bibfield  {author} {\bibinfo {author} {\bibfnamefont {B.~F.}\ \bibnamefont
  {Schutz}},\ }\bibfield  {title} {\bibinfo {title} {{Networks of gravitational
  wave detectors and three figures of merit}},\ }\href
  {https://doi.org/10.1088/0264-9381/28/12/125023} {\bibfield  {journal}
  {\bibinfo  {journal} {Class. Quant. Grav.}\ }\textbf {\bibinfo {volume}
  {28}},\ \bibinfo {pages} {125023} (\bibinfo {year} {2011})},\ \Eprint
  {https://arxiv.org/abs/1102.5421} {arXiv:1102.5421 [astro-ph.IM]}
  \BibitemShut {NoStop}%
\bibitem [{\citenamefont {Abbott}\ \emph {et~al.}(2024)\citenamefont {Abbott}
  \emph {et~al.}}]{LIGOScientific:2021usb}%
  \BibitemOpen
  \bibfield  {author} {\bibinfo {author} {\bibfnamefont {R.}~\bibnamefont
  {Abbott}} \emph {et~al.} (\bibinfo {collaboration} {LIGO Scientific,
  VIRGO}),\ }\bibfield  {title} {\bibinfo {title} {{GWTC-2.1: Deep extended
  catalog of compact binary coalescences observed by LIGO and Virgo during the
  first half of the third observing run}},\ }\href
  {https://doi.org/10.1103/PhysRevD.109.022001} {\bibfield  {journal} {\bibinfo
   {journal} {Phys. Rev. D}\ }\textbf {\bibinfo {volume} {109}},\ \bibinfo
  {pages} {022001} (\bibinfo {year} {2024})},\ \Eprint
  {https://arxiv.org/abs/2108.01045} {arXiv:2108.01045 [gr-qc]} \BibitemShut
  {NoStop}%
\bibitem [{\citenamefont {Dent}(2023)}]{o4_live_pastro_techdoc}%
  \BibitemOpen
  \bibfield  {author} {\bibinfo {author} {\bibfnamefont {T.}~\bibnamefont
  {Dent}},\ }\href {{https://dcc.ligo.org/LIGO-T2300168/public}} {\bibinfo
  {title} {{Technical note: PyCBC Live p\_astro for O4}}},\ \bibinfo
  {howpublished} {DCC LIGO-T2300168-v2} (\bibinfo {year} {2023})\BibitemShut
  {NoStop}%
\bibitem [{\citenamefont {Villa-Ortega}\ \emph {et~al.}(2022)\citenamefont
  {Villa-Ortega}, \citenamefont {Dent},\ and\ \citenamefont
  {Barroso}}]{Villa-Ortega:2022qdo}%
  \BibitemOpen
  \bibfield  {author} {\bibinfo {author} {\bibfnamefont {V.}~\bibnamefont
  {Villa-Ortega}}, \bibinfo {author} {\bibfnamefont {T.}~\bibnamefont {Dent}},\
  and\ \bibinfo {author} {\bibfnamefont {A.~C.}\ \bibnamefont {Barroso}},\
  }\bibfield  {title} {\bibinfo {title} {{Rapid source classification and
  distance estimation for compact binary mergers with PyCBC live}},\ }\href
  {https://doi.org/10.1093/mnras/stac2120} {\bibfield  {journal} {\bibinfo
  {journal} {Mon. Not. Roy. Astron. Soc.}\ }\textbf {\bibinfo {volume} {515}},\
  \bibinfo {pages} {5718} (\bibinfo {year} {2022})},\ \Eprint
  {https://arxiv.org/abs/2203.10080} {arXiv:2203.10080 [astro-ph.HE]}
  \BibitemShut {NoStop}%
\bibitem [{\citenamefont {{LVK}}(2026)}]{userguide-inference}%
  \BibitemOpen
  \bibfield  {author} {\bibinfo {author} {\bibnamefont {{LVK}}},\ }\href@noop
  {} {\bibinfo {title} {{IGWN Public Alerts User Guide}}},\ \bibinfo
  {howpublished}
  {{https://emfollow.docs.ligo.org/userguide/content.html\#inference}}
  (\bibinfo {year} {2026})\BibitemShut {NoStop}%
\bibitem [{\citenamefont {Metzger}(2020)}]{Metzger:2019zeh}%
  \BibitemOpen
  \bibfield  {author} {\bibinfo {author} {\bibfnamefont {B.~D.}\ \bibnamefont
  {Metzger}},\ }\bibfield  {title} {\bibinfo {title} {{Kilonovae}},\ }\href
  {https://doi.org/10.1007/s41114-019-0024-0} {\bibfield  {journal} {\bibinfo
  {journal} {Living Rev. Rel.}\ }\textbf {\bibinfo {volume} {23}},\ \bibinfo
  {pages} {1} (\bibinfo {year} {2020})},\ \Eprint
  {https://arxiv.org/abs/1910.01617} {arXiv:1910.01617 [astro-ph.HE]}
  \BibitemShut {NoStop}%
\bibitem [{\citenamefont {Burns}(2020)}]{Burns:2019byj}%
  \BibitemOpen
  \bibfield  {author} {\bibinfo {author} {\bibfnamefont {E.}~\bibnamefont
  {Burns}},\ }\bibfield  {title} {\bibinfo {title} {{Neutron Star Mergers and
  How to Study Them}},\ }\href {https://doi.org/10.1007/s41114-020-00028-7}
  {\bibfield  {journal} {\bibinfo  {journal} {Living Rev. Rel.}\ }\textbf
  {\bibinfo {volume} {23}},\ \bibinfo {pages} {4} (\bibinfo {year} {2020})},\
  \Eprint {https://arxiv.org/abs/1909.06085} {arXiv:1909.06085 [astro-ph.HE]}
  \BibitemShut {NoStop}%
\bibitem [{\citenamefont {{Nitz}}\ \emph {et~al.}(2020)\citenamefont {{Nitz}},
  \citenamefont {{Sch{\"a}fer}},\ and\ \citenamefont {{Dal
  Canton}}}]{MergerForecasting}%
  \BibitemOpen
  \bibfield  {author} {\bibinfo {author} {\bibfnamefont {A.~H.}\ \bibnamefont
  {{Nitz}}}, \bibinfo {author} {\bibfnamefont {M.}~\bibnamefont
  {{Sch{\"a}fer}}},\ and\ \bibinfo {author} {\bibfnamefont {T.}~\bibnamefont
  {{Dal Canton}}},\ }\bibfield  {title} {\bibinfo {title} {{Gravitational-wave
  Merger Forecasting: Scenarios for the Early Detection and Localization of
  Compact-binary Mergers with Ground-based Observatories}},\ }\href
  {https://doi.org/10.3847/2041-8213/abbc10} {\bibfield  {journal} {\bibinfo
  {journal} {\apjl}\ }\textbf {\bibinfo {volume} {902}},\ \bibinfo {eid} {L29}
  (\bibinfo {year} {2020})},\ \Eprint {https://arxiv.org/abs/2009.04439}
  {arXiv:2009.04439 [astro-ph.HE]} \BibitemShut {NoStop}%
\bibitem [{\citenamefont {{Tohuvavohu}}\ \emph {et~al.}(2024)\citenamefont
  {{Tohuvavohu}}, \citenamefont {{Kennea}}, \citenamefont {{Roberts}},
  \citenamefont {{DeLaunay}}, \citenamefont {{Ronchini}}, \citenamefont
  {{Cenko}}, \citenamefont {{Ewing}}, \citenamefont {{Magee}}, \citenamefont
  {{Messick}}, \citenamefont {{Sachdev}},\ and\ \citenamefont
  {{Singer}}}]{SwiftlyChasing}%
  \BibitemOpen
  \bibfield  {author} {\bibinfo {author} {\bibfnamefont {A.}~\bibnamefont
  {{Tohuvavohu}}}, \bibinfo {author} {\bibfnamefont {J.~A.}\ \bibnamefont
  {{Kennea}}}, \bibinfo {author} {\bibfnamefont {C.~J.}\ \bibnamefont
  {{Roberts}}}, \bibinfo {author} {\bibfnamefont {J.}~\bibnamefont
  {{DeLaunay}}}, \bibinfo {author} {\bibfnamefont {S.}~\bibnamefont
  {{Ronchini}}}, \bibinfo {author} {\bibfnamefont {S.~B.}\ \bibnamefont
  {{Cenko}}}, \bibinfo {author} {\bibfnamefont {B.}~\bibnamefont {{Ewing}}},
  \bibinfo {author} {\bibfnamefont {R.}~\bibnamefont {{Magee}}}, \bibinfo
  {author} {\bibfnamefont {C.}~\bibnamefont {{Messick}}}, \bibinfo {author}
  {\bibfnamefont {S.}~\bibnamefont {{Sachdev}}},\ and\ \bibinfo {author}
  {\bibfnamefont {L.~P.}\ \bibnamefont {{Singer}}},\ }\bibfield  {title}
  {\bibinfo {title} {{Swiftly Chasing Gravitational Waves across the Sky in
  Real Time}},\ }\href {https://doi.org/10.3847/2041-8213/ad87ce} {\bibfield
  {journal} {\bibinfo  {journal} {\apjl}\ }\textbf {\bibinfo {volume} {975}},\
  \bibinfo {eid} {L19} (\bibinfo {year} {2024})},\ \Eprint
  {https://arxiv.org/abs/2410.05720} {arXiv:2410.05720 [astro-ph.HE]}
  \BibitemShut {NoStop}%
\bibitem [{\citenamefont {Brown}\ \emph {et~al.}(2012)\citenamefont {Brown},
  \citenamefont {Harry}, \citenamefont {Lundgren},\ and\ \citenamefont
  {Nitz}}]{Brown:2012qf}%
  \BibitemOpen
  \bibfield  {author} {\bibinfo {author} {\bibfnamefont {D.~A.}\ \bibnamefont
  {Brown}}, \bibinfo {author} {\bibfnamefont {I.}~\bibnamefont {Harry}},
  \bibinfo {author} {\bibfnamefont {A.}~\bibnamefont {Lundgren}},\ and\
  \bibinfo {author} {\bibfnamefont {A.~H.}\ \bibnamefont {Nitz}},\ }\bibfield
  {title} {\bibinfo {title} {{Detecting binary neutron star systems with spin
  in advanced gravitational-wave detectors}},\ }\href
  {https://doi.org/10.1103/PhysRevD.86.084017} {\bibfield  {journal} {\bibinfo
  {journal} {Phys. Rev. D}\ }\textbf {\bibinfo {volume} {86}},\ \bibinfo
  {pages} {084017} (\bibinfo {year} {2012})},\ \Eprint
  {https://arxiv.org/abs/1207.6406} {arXiv:1207.6406 [gr-qc]} \BibitemShut
  {NoStop}%
\bibitem [{\citenamefont {{LIGO Scientific
  Collaboration}}(2010)}]{aLIGODesign}%
  \BibitemOpen
  \bibfield  {author} {\bibinfo {author} {\bibnamefont {{LIGO Scientific
  Collaboration}}},\ }\href {https://dcc.ligo.org/LIGO-T0900288/public}
  {\bibinfo {title} {{Advanced {LIGO} anticipated sensitivities}}},\ \bibinfo
  {howpublished} {LIGO Document T0900288-v3} (\bibinfo {year}
  {2010})\BibitemShut {NoStop}%
\bibitem [{\citenamefont {Tolley}(2024)}]{Tolley:2025}%
  \BibitemOpen
  \bibfield  {author} {\bibinfo {author} {\bibfnamefont {A.}~\bibnamefont
  {Tolley}},\ }\emph {\bibinfo {title} {{Improving the Detection of
  Gravitational-Wave Signals in Real Time}}},\ \href@noop {} {Ph.D. thesis},\
  \bibinfo  {school} {University of Portsmouth} (\bibinfo {year} {2024}),\
  \Eprint {https://arxiv.org/abs/2503.21417} {arXiv:2503.21417 [gr-qc]}
  \BibitemShut {NoStop}%
\bibitem [{\citenamefont {Abbott}\ \emph
  {et~al.}(2017{\natexlab{b}})\citenamefont {Abbott} \emph
  {et~al.}}]{GW170817}%
  \BibitemOpen
  \bibfield  {author} {\bibinfo {author} {\bibfnamefont {B.~P.}\ \bibnamefont
  {Abbott}} \emph {et~al.} (\bibinfo {collaboration} {LIGO Scientific,
  Virgo}),\ }\bibfield  {title} {\bibinfo {title} {{GW170817: Observation of
  Gravitational Waves from a Binary Neutron Star Inspiral}},\ }\href
  {https://doi.org/10.1103/PhysRevLett.119.161101} {\bibfield  {journal}
  {\bibinfo  {journal} {Phys. Rev. Lett.}\ }\textbf {\bibinfo {volume} {119}},\
  \bibinfo {pages} {161101} (\bibinfo {year} {2017}{\natexlab{b}})},\ \Eprint
  {https://arxiv.org/abs/1710.05832} {arXiv:1710.05832 [gr-qc]} \BibitemShut
  {NoStop}%
\bibitem [{\citenamefont {Storn}\ and\ \citenamefont
  {Price}(1997)}]{Storn:1997}%
  \BibitemOpen
  \bibfield  {author} {\bibinfo {author} {\bibfnamefont {R.}~\bibnamefont
  {Storn}}\ and\ \bibinfo {author} {\bibfnamefont {K.}~\bibnamefont {Price}},\
  }\bibfield  {title} {\bibinfo {title} {{Differential Evolution -- A Simple
  and Efficient Heuristic for Global Optimization over Continuous Spaces}},\
  }\href {https://doi.org/10.1023/A:1008202821328} {\bibfield  {journal}
  {\bibinfo  {journal} {Journal of Global Optimization}\ }\textbf {\bibinfo
  {volume} {11}},\ \bibinfo {pages} {341} (\bibinfo {year} {1997})}\BibitemShut
  {NoStop}%
\bibitem [{\citenamefont {Virtanen}\ \emph {et~al.}(2020)\citenamefont
  {Virtanen} \emph {et~al.}}]{Virtanen:2019joe}%
  \BibitemOpen
  \bibfield  {author} {\bibinfo {author} {\bibfnamefont {P.}~\bibnamefont
  {Virtanen}} \emph {et~al.},\ }\bibfield  {title} {\bibinfo {title} {{SciPy
  1.0--Fundamental Algorithms for Scientific Computing in Python}},\ }\href
  {https://doi.org/10.1038/s41592-019-0686-2} {\bibfield  {journal} {\bibinfo
  {journal} {Nat. Meth.}\ }\textbf {\bibinfo {volume} {17}},\ \bibinfo {pages}
  {261} (\bibinfo {year} {2020})},\ \Eprint {https://arxiv.org/abs/1907.10121}
  {arXiv:1907.10121 [cs.MS]} \BibitemShut {NoStop}%
\bibitem [{\citenamefont {Chaudhary}\ \emph {et~al.}(2024)\citenamefont
  {Chaudhary} \emph {et~al.}}]{Chaudhary:2023}%
  \BibitemOpen
  \bibfield  {author} {\bibinfo {author} {\bibfnamefont {S.~S.}\ \bibnamefont
  {Chaudhary}} \emph {et~al.},\ }\bibfield  {title} {\bibinfo {title}
  {{Low-latency gravitational wave alert products and their performance at the
  time of the fourth LIGO-Virgo-KAGRA observing run}},\ }\href
  {https://doi.org/10.1073/pnas.2316474121} {\bibfield  {journal} {\bibinfo
  {journal} {Proc. Nat. Acad. Sci.}\ }\textbf {\bibinfo {volume} {121}},\
  \bibinfo {pages} {e2316474121} (\bibinfo {year} {2024})},\ \Eprint
  {https://arxiv.org/abs/2308.04545} {arXiv:2308.04545 [astro-ph.HE]}
  \BibitemShut {NoStop}%
\bibitem [{\citenamefont {Hannam}\ \emph {et~al.}(2014)\citenamefont {Hannam},
  \citenamefont {Schmidt}, \citenamefont {Boh\'e}, \citenamefont {Haegel},
  \citenamefont {Husa}, \citenamefont {Ohme}, \citenamefont {Pratten},\ and\
  \citenamefont {P\"urrer}}]{Hannam:2013oca}%
  \BibitemOpen
  \bibfield  {author} {\bibinfo {author} {\bibfnamefont {M.}~\bibnamefont
  {Hannam}}, \bibinfo {author} {\bibfnamefont {P.}~\bibnamefont {Schmidt}},
  \bibinfo {author} {\bibfnamefont {A.}~\bibnamefont {Boh\'e}}, \bibinfo
  {author} {\bibfnamefont {L.}~\bibnamefont {Haegel}}, \bibinfo {author}
  {\bibfnamefont {S.}~\bibnamefont {Husa}}, \bibinfo {author} {\bibfnamefont
  {F.}~\bibnamefont {Ohme}}, \bibinfo {author} {\bibfnamefont {G.}~\bibnamefont
  {Pratten}},\ and\ \bibinfo {author} {\bibfnamefont {M.}~\bibnamefont
  {P\"urrer}},\ }\bibfield  {title} {\bibinfo {title} {{Simple Model of
  Complete Precessing Black-Hole-Binary Gravitational Waveforms}},\ }\href
  {https://doi.org/10.1103/PhysRevLett.113.151101} {\bibfield  {journal}
  {\bibinfo  {journal} {Phys. Rev. Lett.}\ }\textbf {\bibinfo {volume} {113}},\
  \bibinfo {pages} {151101} (\bibinfo {year} {2014})},\ \Eprint
  {https://arxiv.org/abs/1308.3271} {arXiv:1308.3271 [gr-qc]} \BibitemShut
  {NoStop}%
\bibitem [{\citenamefont {Dietrich}\ \emph {et~al.}(2019)\citenamefont
  {Dietrich}, \citenamefont {Samajdar}, \citenamefont {Khan}, \citenamefont
  {Johnson-McDaniel}, \citenamefont {Dudi},\ and\ \citenamefont
  {Tichy}}]{Dietrich:2019kaq}%
  \BibitemOpen
  \bibfield  {author} {\bibinfo {author} {\bibfnamefont {T.}~\bibnamefont
  {Dietrich}}, \bibinfo {author} {\bibfnamefont {A.}~\bibnamefont {Samajdar}},
  \bibinfo {author} {\bibfnamefont {S.}~\bibnamefont {Khan}}, \bibinfo {author}
  {\bibfnamefont {N.~K.}\ \bibnamefont {Johnson-McDaniel}}, \bibinfo {author}
  {\bibfnamefont {R.}~\bibnamefont {Dudi}},\ and\ \bibinfo {author}
  {\bibfnamefont {W.}~\bibnamefont {Tichy}},\ }\bibfield  {title} {\bibinfo
  {title} {{Improving the NRTidal model for binary neutron star systems}},\
  }\href {https://doi.org/10.1103/PhysRevD.100.044003} {\bibfield  {journal}
  {\bibinfo  {journal} {Phys. Rev. D}\ }\textbf {\bibinfo {volume} {100}},\
  \bibinfo {pages} {044003} (\bibinfo {year} {2019})},\ \Eprint
  {https://arxiv.org/abs/1905.06011} {arXiv:1905.06011 [gr-qc]} \BibitemShut
  {NoStop}%
\bibitem [{\citenamefont {Douchin}\ and\ \citenamefont
  {Haensel}(2001)}]{Douchin:2001sv}%
  \BibitemOpen
  \bibfield  {author} {\bibinfo {author} {\bibfnamefont {F.}~\bibnamefont
  {Douchin}}\ and\ \bibinfo {author} {\bibfnamefont {P.}~\bibnamefont
  {Haensel}},\ }\bibfield  {title} {\bibinfo {title} {{A unified equation of
  state of dense matter and neutron star structure}},\ }\href
  {https://doi.org/10.1051/0004-6361:20011402} {\bibfield  {journal} {\bibinfo
  {journal} {Astron. Astrophys.}\ }\textbf {\bibinfo {volume} {380}},\ \bibinfo
  {pages} {151} (\bibinfo {year} {2001})},\ \Eprint
  {https://arxiv.org/abs/astro-ph/0111092} {arXiv:astro-ph/0111092}
  \BibitemShut {NoStop}%
\bibitem [{\citenamefont {Aghanim}\ \emph {et~al.}(2020)\citenamefont {Aghanim}
  \emph {et~al.}}]{Planck:2018vyg}%
  \BibitemOpen
  \bibfield  {author} {\bibinfo {author} {\bibfnamefont {N.}~\bibnamefont
  {Aghanim}} \emph {et~al.} (\bibinfo {collaboration} {Planck}),\ }\bibfield
  {title} {\bibinfo {title} {{Planck 2018 results. VI. Cosmological
  parameters}},\ }\href {https://doi.org/10.1051/0004-6361/201833910}
  {\bibfield  {journal} {\bibinfo  {journal} {Astron. Astrophys.}\ }\textbf
  {\bibinfo {volume} {641}},\ \bibinfo {pages} {A6} (\bibinfo {year} {2020})},\
  \bibinfo {note} {[Erratum: Astron.Astrophys. 652, C4 (2021)]},\ \Eprint
  {https://arxiv.org/abs/1807.06209} {arXiv:1807.06209 [astro-ph.CO]}
  \BibitemShut {NoStop}%
\bibitem [{\citenamefont {Insley}\ \emph {et~al.}(2026)\citenamefont {Insley},
  \citenamefont {Williams}, \citenamefont {Dhurkunde},\ and\ \citenamefont
  {Harry}}]{Insley:2026tpg}%
  \BibitemOpen
  \bibfield  {author} {\bibinfo {author} {\bibfnamefont {S.}~\bibnamefont
  {Insley}}, \bibinfo {author} {\bibfnamefont {M.~J.}\ \bibnamefont
  {Williams}}, \bibinfo {author} {\bibfnamefont {R.}~\bibnamefont
  {Dhurkunde}},\ and\ \bibinfo {author} {\bibfnamefont {I.}~\bibnamefont
  {Harry}},\ }\bibfield  {title} {\bibinfo {title} {{Normalizing flows for
  density estimation in multi-detector gravitational-wave searches}},\
  }\href@noop {} {\bibfield  {journal} {\bibinfo  {journal} {arXiv}\ }
  (\bibinfo {year} {2026})},\ \Eprint {https://arxiv.org/abs/2604.26581}
  {arXiv:2604.26581 [astro-ph.HE]} \BibitemShut {NoStop}%
\bibitem [{\citenamefont {Nitz}\ \emph {et~al.}(2026)\citenamefont {Nitz} \emph
  {et~al.}}]{pycbc-software}%
  \BibitemOpen
  \bibfield  {author} {\bibinfo {author} {\bibfnamefont {A.}~\bibnamefont
  {Nitz}} \emph {et~al.},\ }\href@noop {} {\bibinfo {title} {{PyCBC: A Python
  toolkit for gravitational-wave astronomy}}},\ \bibinfo {howpublished}
  {\url{https://pycbc.org/}} (\bibinfo {year} {2026})\BibitemShut {NoStop}%
\end{thebibliography}%
\end{document}